\documentclass[12pt]{iopart}

\usepackage{graphicx}
\usepackage{iopams}

\newcommand \be{\begin{eqnarray}}
\newcommand \ee{\end{eqnarray}}
\newcommand \ba{\begin{eqnarray}}
\newcommand \ea{\end{eqnarray}}

\begin{document}

\title[Positive streamers at varying density: experiments on similarity laws]
{Positive streamers in air and nitrogen of varying density: experiments on similarity laws}

\author{T M P Briels$^1$, E M van Veldhuizen$^1$ and U Ebert$^{1,2}$}

\address{$^1$ Department of Applied Physics, Technische Universiteit
Eindhoven, P O Box 513, 5600 MB Eindhoven, The Netherlands,\\
$^2$ Centrum voor Wiskunde en Informatica (CWI), P O Box 94079, 1090 GB Amsterdam, The Netherlands}

\ead{ebert@cwi.nl, e.m.v.veldhuizen@tue.nl}

\begin{abstract}
Positive streamers in ambient air at pressures from 0.013 to 1 bar are investigated experimentally. The
voltage applied to the anode needle ranges from 5 to 45 kV, the discharge gap from 1 to 16 cm. Using a
"slow" voltage rise time of 100 to 180 ns, the streamers are intentionally kept thin. For each pressure $p$,
we find a minimal diameter $d_{min}$. To test whether streamers at different pressures are similar, the
minimal streamer diameter $d_{\rm min}$ is multiplied by its pressure $p$; we find this product to be well
approximated by $p\cdot d_{\rm min}=0.20\pm 0.02~{\rm mm\cdot bar}$ over two decades of air pressure at room
temperature. The value also fits diameters of sprite discharges above thunderclouds at an altitude of 80 km
when extrapolated to room temperature (as air density rather than pressure determines the physical
behavior). The minimal velocity of streamers in our measurements is approximately 0.1 mm/ns = $10^5$ m/s.
The same minimal velocity has been reported for tendrils in sprites. We also investigate the size of the
initial ionization cloud at the electrode tip from which the streamers emerge, and the streamer length
between branching events. The same quantities are also measured in nitrogen with a purity of approximately
99.9\%. We characterize the essential differences with streamers in air and find a minimal diameter of
$p\cdot d_{\rm min}=0.12\pm 0.02~{\rm mm\cdot bar}$ in our nitrogen.
\end{abstract}


\submitto{The paper is accepted for the cluster issue on ``Streamers, Sprites and Lightning'' in \JPD}

\maketitle

\section{Introduction}

\subsection{Streamers in different media at various densities}

Streamer discharges determine the early stages of sparks and lightning; they occur in various ionizable
media in a large range of pressures and temperatures. In corona applications, they are widely used for
surface charging in copiers and printers, for the removal of dust in electric precipitators and for ozone
production and gas and water cleaning~\cite{eddieboek,keping,pok,ebe06,HansW,LC,Cluster}. Streamers underlie
the operation of spark plugs in internal combustion engines, and there are new initiatives to improve the
control of ignition~\cite{Renault,Stara}. Another new application is flow control in
aviation~\cite{StariW,Pitch}. Streamers also play a role in the ignition of energy efficient high pressure
metal halid lamps~\cite{Philips}. Obviously, pressure, temperature and gas composition vary from one
phenomenon to the next. In electrical engineering, the higher densities of liquids are desirable for fast
switching, but pre-existing microbubbles largely influence streamer properties in fluids~\cite{Schoen,An}.
For this reason, the operation of spark gaps filled with supercritical fluids is now under
investigation~\cite{Bert}. We also recall experimental work on streamers in fluid nitrogen and argon
~\cite{Denat}. Streamer-like phenomena also occur in fast semiconductor switches~\cite{Rodin}. While in all
these cases, the density of neutral particles is of the order of or considerably larger than in air at
standard temperature and pressure, much lower densities are of interest in geophysics. At altitudes of 40-90
km in the atmosphere, enormous sprite discharges~\cite{6,7,gerken,McHarg,Nielsen07} were found to evolve
above thunderclouds. They are thought to be related to streamer discharges through similarity
laws~\cite{ebe06,pasko2007} as elaborated in the next section.

\subsection{Townsend scaling: theory of similarity laws for varying density\label{Townsend}}

\begin{figure}
\begin{center}
\vspace{0.3cm}
\includegraphics[width=16cm]{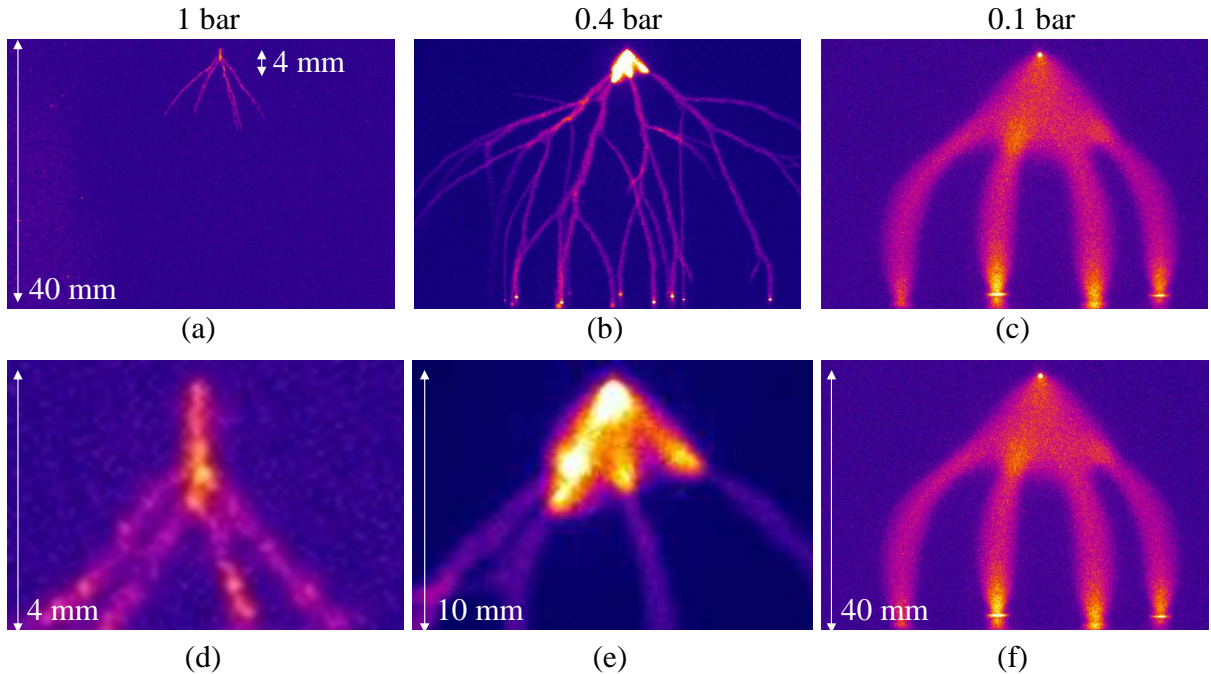}
\caption{Visualization of similarity. Positive streamers in a 4 cm gap in air at a voltage of 10 kV at room
temperature and varying pressure: 1 bar (a, d), 0.4 bar (b, e), and 0.1 bar (c, f). Panels a to c show the
whole 4 cm gap, panels d to f zoom into the upper panels in such a way that panel height times pressure are
always 4 mm$\cdot$bar. Number and diameters of streamers are similar in the lower panels at some distance
from the needle electrode. The pictures show the complete streamer evolution, hence the streamers in panel
(a) do not reach the plate electrode. Note also, that in the extreme zoom of panel (d), the pixel structure
of the camera becomes visible and diameter evaluations become inaccurate. The diameters reported in this
paper are evaluated from pictures where the camera is moved physically so close to the discharge that such
artifacts are avoided~\cite{bri06}.} \label{drukzoom}
\end{center}
\end{figure}

Similarity laws between streamer discharges at different densities $n$ of the neutral gas or other medium
follow from the fact that the basic length scale of the streamer discharge is the mean free path $\ell_{\rm
MFP}$ of the electron which is inversely proportional to the density of the medium: $\ell_{\rm MFP}\propto
1/n$. An electron is accelerated over this length by the local electric field {\bf E} before the next
collision with a neutral particle; on this length it gains the energy ${\rm e}\;|{\bf E}|\;\ell_{\rm MFP}$
where e is the elementary charge; this energy has to be compared to the ionization energy of the neutral
particle on which the electron impacts. Similarity for varying density $n$ therefore implies that the reduced
electric field ${\bf E}/n$ is the same at similar places, that the streamer velocity $v$ as well as velocity
and energy distributions of individual electrons are the same at similar places, that all reduced lengths,
i.e., the ratio of lengths $\ell$ over the mean free electron paths $\ell/\ell_{\rm MFP}\propto \ell\cdot n$
are the same, and that the same electric potential is applied over the same reduced lengths --- these
similarities have motivated the introduction of the unit Townsend for the ratio ${\bf E}/n$ of electric field
over density, and we therefore will denote these similarity laws as Townsend scaling. The similarity can be
tested on the experimental results as demonstrated in Fig.~\ref{drukzoom}, for a detailed description of
these experiments, we refer to later chapters.

A minimal model~\cite{ebe06,UWCPRE97} for negative streamers with drift and diffusion of electrons, local
impact ionization and space charge effects obeys these similarity laws perfectly. The same is true when the
model is extended by nonlocal photo-ionization in the low density limit; this photo-ionization is commonly
thought to explain the propagation of positive streamers in air~\cite{zhel,morrow,kulik}.

Corrections to the similarity laws come from three different sources.\\
$(i)$ The similarity laws are based on the fast two-particle processes between a charged particle and one of
the abundant neutral particles that dominate the dynamics in the streamer ionization front. At higher
densities, three-body processes inside the gas volume can become important. Important examples include
collisional quenching of the excited states that otherwise lead to photo-ionization in nitrogen-oxygen
mixtures like air, as well as electron attachment to oxygen or electron ion
recombination~\cite{pasko2007,zhel,pasko98,Bab2001,PE,alejandro}; in all these cases the conservation of
both momentum and energy requires the presence of a third particle, and the similarity laws break down on
spatial or temporal scales where these processes become important.\\
$(ii)$ Corrections also come from the presence of electrodes or other matter (like dust particles or gas
bubbles) whose lengths do not change when the gas density changes~\cite{Marode,babaeva}.\\
$(iii)$ Finally, the total number of charged particles in similar streamers varies like $1/n$~\cite{ebe06},
therefore similar streamers at higher densities $n$ contain a lower total number of charged particles, and
discrete particle fluctuations are more pronounced.\\
Due to $(i)$ and $(ii)$, similarity laws for positive streamers in air are expected to hold only well below
30 torr, i.e., 40 mbar (1 bar equals $10^5$~Pa), where collisional quenching of photo-ionizing nitrogen
states is negligible~\cite{zhel,alejandro,pasko2006}, in the streamer head when it is far enough from the
electrodes, and at time scales when the loss of streamer conductivity due to electron attachment is still
negligible. We remark that the predicted propagation of positive streamers in absolutely pure nitrogen
depends on the model input about photo-ionization in this gas, but the purity of our nitrogen is only
99.9\%, therefore the common photo-ionization model for nitrogen-oxygen mixtures is still likely to apply.

The basic physical similarity variable is the density $n$ of neutral particles, therefore reduced fields and
lengths should be defined as ${\bf E}/n$ and $\ell\cdot n$. The density $n$ is related to the pressure $p$
approximately as $n=p/(k_BT)$ according to the ideal gas law, where $k_B$ is the Boltzmann constant and $T$
is temperature. As our experiments are all performed at room temperature and as we measure pressure directly,
we discuss the pressure dependence of our laboratory experiments rather than their density dependence and use
${\bf E}/p$ and $\ell\cdot p$. We extrapolate sprite measurements to room temperature using the ideal gas
law.

\subsection{Previous experimental investigations}

Experimental investigations of the density dependence of streamers are rare. Next to own preliminary
visualizing work~\cite{bri05}, Pancheshnyi {\it et al.}~\cite{panchpre} have varied the pressure in the
range of 0.5 to 1 bar in a 2 cm point-plane gap in synthetic air at a fixed voltage of 24 kV; they find that
the streamer diameter increases from 0.4 mm to 2.6 mm with decreasing pressure. Becerra {\it et
al.}~\cite{Bec} have presented yet unpublished results in the pressure range of 0.01 to 0.06 bar at a recent
workshop~\cite{LC}. Streamers in air in a DC corona at pressures of 1-10 bar were observed by Marode {\it et
al.}~\cite{Marode}. They conclude that the general discharge structures and, in particular, the streamer
velocities, are similar when pressure changes.

In sprite discharges, channels have been imaged with a telescope~\cite{gerken}. Channels of varying diameter
at fixed height are reported, e.g., 60 to 145 m ($\pm$ 12 m) in the altitude range of 60-64 km ($\pm$ 4.5
km) and 20-50 m ($\pm$ 13 m) in the altitude range of 76-80 km ($\pm$ 5 km)~\cite{gerken}. The variation of
sprite diameters resembles the variation of streamer diameters at 1 bar~\cite{bri06,bri08jpd,luque08jpd}.
The sprites extend downward with velocities that vary by more than two orders of magnitude, ranging from 0.1
mm/ns (= $10^5$ m/s) to more than 3$\cdot 10^7$ m/s \cite{moudry} or even $10^8$ m/s~\cite{McHarg02}.

Other experimental work is concerned with the influence of voltage and gas composition at standard
temperature and pressure on positive streamers. In previous and forth coming
work~\cite{bri06,bri08jpd,luque08jpd}, we have shown that diameter and velocity of positive streamers in air
at standard temperature and pressure in point-plane gaps of 4 and 8 cm can vary by more than an order of
magnitude as a function of voltage, i.e., from 0.2 to 3 mm and from 0.1 to 4 mm/ns for voltages between 5
and 96 kV; furthermore, the diameters and velocities depend not only on the maximal voltage, but also on
voltage rise time and possible current limitations~\cite{bri06}. A range of diameters is also presented by
Ono {\it et al.}~\cite{ono} where streamer diameters increase from 0.4 mm to 1.2 mm in a 1.3 cm point-plane
gap in synthetic air at 1 bar when the voltage increases from 13 kV to 27 kV. The average streamer velocity
is 0.5 mm/ns at 18 kV. In nitrogen they find the diameter to increase from 0.2 mm to 0.4 mm and the velocity
to increase from 0.2 mm/ns to 0.5 mm/ns when the voltage increases from 16 kV to 27 kV and from 18 kV to 27
kV, respectively. Streamers in a protrusion-plane gap of 13 cm at 1 bar for a range of nitrogen-oxygen
ratios and in the voltage range of 83 to 123 kV are described by Yi and Williams~\cite{yi02}. They observe
an increase in propagation velocity from about 0.3 mm/ns to 4 mm/ns for increasing oxygen concentration (up
to 10\%) and increasing applied voltage.

\subsection{Subject of study: similarity for varying density in air and nitrogen}

Obviously, there is a need to classify these phenomena. Within the present paper, we will investigate
experimentally whether there are similarity laws between positive (cathode directed) streamers in the same
gas type at different densities. We will explore the density regime corresponding to pressures of 0.013 to 1
bar at room temperature and we will discuss whether our results over nearly two decades of density
extrapolate to sprite discharges at much lower densities. If streamers at different densities are similar,
then the huge sprite discharges at very low pressures can be investigated in small laboratory experiments at
much higher pressure, and the fast temporal evolution of streamers in turn can be conveniently observed in
slow motion and under the magnifying glass at lower pressures; for the extreme case of sprites this has been
recently demonstrated in~\cite{McHarg,Nielsen07}.

Most of our measurements have been performed in ambient air as it is most common in applications and in
nature. However, air is a compound gas with 78\% N$_2$, 20\% O$_2$, 1\% H$_2$O, 0.03\% CO$_2$ and 1\% noble
gases; processes like photo-ionization between nitrogen and oxygen and electron attachment to oxygen have to
be taken into account. For this reason also experiments in nitrogen as a simple gas are performed and
presented; however, we here only present results at a nitrogen purity of about 99.9~\%, where impurities
still seem to play a role.

This paper concentrates on measurements of streamer diameters, and also reports the size of the initial
ionization cloud at inception and the  velocities, and it presents first results on branching lengths.
Branching angles were recently measured in~\cite{Nijdam08}. As shown in~\cite{bri06} and discussed in more
detail in~\cite{bri08jpd}, there is a critical inception voltage required for streamer formation; at the
inception voltage, streamers at a given pressure always appear to attain the same minimal diameter. When a
higher voltage is applied within a sufficiently short rise time, considerably thicker and faster streamers
are formed~\cite{bri06,bri08jpd}. However, when the voltage is increased sufficiently slowly, streamers
emerge as soon as the inception voltage is reached, and they (obviously) have minimal diameter. In the
present paper, we have chosen to test the similarity laws on these thinnest streamers for several reasons.
First, it appears that there is a lower limit for the streamer diameter but an upper limit is not
established \cite{bri06,bri08jpd}. Therefore, the minimal diameters at each pressure are appropriate
quantities to compare. Second, the thinnest streamers are furthest away from the electrodes that might break
the similarity laws locally as discussed in section~\ref{Townsend}. Third, the minimal streamers do not
evolve into glow and/or spark as easily as the thick ones, and therefore they can be imaged more easily.

The paper is organized as follows. Section 2 contains a detailed description of the experimental setup,
section 3 presents the measurements of streamer diameters and of the size of the initial cloud, of streamer
velocities and of branching ratios. Section 4 contains discussion, conclusions and outlook. Appendix A
illustrates the difference between streamers in air and in our N$_2$-O$_2$ mixture.


\section{Experimental details}\label{setup}

\subsection{Experimental setup}

\begin{figure}
\begin{center}
\vspace{.5cm}
\includegraphics[width=16cm]{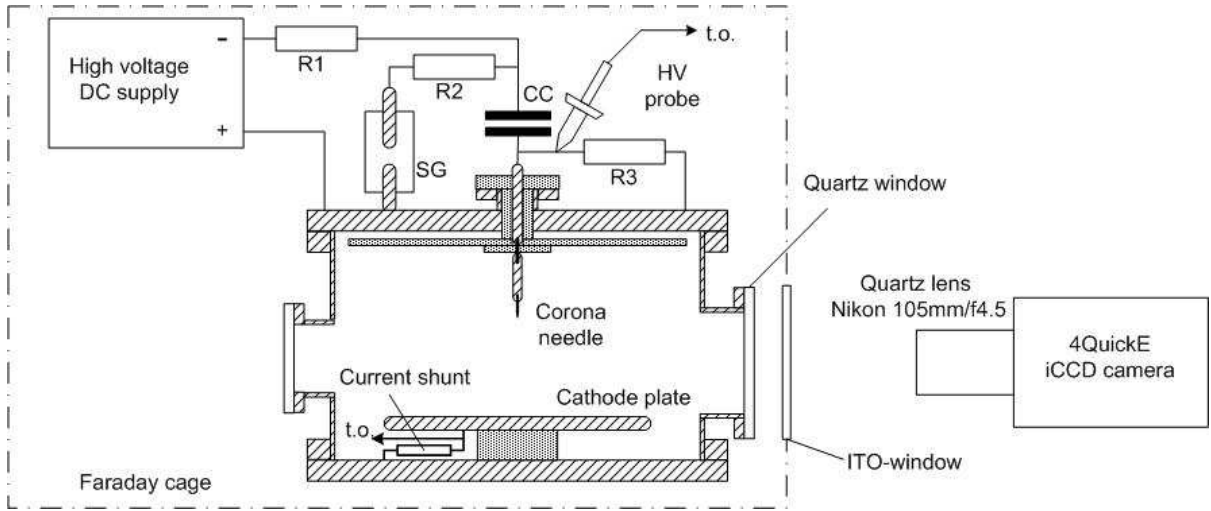}
\caption{Scheme of the stainless steel vacuum vessel with the sparkgap C-supply for a 16 cm electrode gap.
For other gap lengths a divided cathode is used \cite{bri06}. SG = spark gap, CC = coupling capacitor, $R_1$
= charging resistor, $R_2$ = rise time determining resistor, $R_3$ = decay time determining resistor, t.o.=
to oscilloscope.} \label{expsetupoverview}
\end{center}
\end{figure}

\begin{table}
\begin{center}
\centering
\begin{tabular}{|l||l|l|l|l|l|l|l|}\hline
gap (cm) & $R_1$  & $R_2$ & $R_3$ & shunt & CC & $t_{\rm rise}$ & switch\\
\hline \hline
1, 2, 4  & 25 M$\Omega$ & 2 k$\Omega$        & 1 M$\Omega$  & 2.75 $\Omega$ & 1 nF & 150-180 ns & SC\\
\hline
16 & 25 M$\Omega$ & 1 or 2 k$\Omega$ & 8 k$\Omega$, 1 or 25 M$\Omega$ & 8.25 $\Omega$  & 1 nF & 100-170 ns & SG\\
\hline
\end{tabular}
\caption{Values of the components used in the C-supply. For $R_1$, $R_2$, $R_3$, shunt and CC, see figure
\ref{expsetupoverview}, $t_{\rm rise}$ = voltage rise time, SC = Behlke HTS 651 semiconductor switch; SG =
homemade sparkgap.} \label{n2airtabexpopzet}
\end{center}
\end{table}

The setup is sketched in Fig.~\ref{expsetupoverview}. Positive streamers are created in a large cylindrical
stainless steel vacuum vessel with an internal diameter of 50 cm and an internal height of 30 cm. The vacuum
vessel has three viewing ports. The viewing port for the camera has a quartz window that is transparent also
for the UV-light of the streamer discharge. The camera is shielded from the setup by a Faraday cage. This
Faraday cage contains a window coated with a conducting layer of Indium-Tin-Oxide (ITO) to keep the cage
intact; it is transparent for wave lengths larger than 300 nm. High voltage pulses of up to 50 kV in gaps up
to 16 cm are generated with the C-supply, that is thoroughly described in \cite{bri06}. The components of
the supply are given in Table~\ref{n2airtabexpopzet}. A relatively large series resistance $R_2$ is used
here to obtain mainly minimal streamers (i.e. type 3 streamers with a diameter of 0.2 mm at 1 bar
\cite{bri06}); the resistor increases the voltage rise time, such that the streamers start before the
voltage has reached its maximal value; and it also limits the current through the discharging
circuit~\cite{bri06}. (Note that a voltage rise time of 30 ns or less and $R_2=0$ is required to create
thick streamers with a diameter of more than 1 mm.)

The times $T$ and $\Delta t$ indicated in the figure captions are defined as follows. $\Delta t$ is the
exposure time of the camera. The absolute time $T=0$ indicates the pulse of the function generator. Now it
should be noted that the spark gap (SG) and the Behlke switch (SC) use different trigger electronics.
Therefore the time delay between the pulse of the function generator that sends a signal to the high voltage
switch and the actual start of the high voltage pulse over the gap is at least 1.2 $\mu$s for the spark gap
\cite{bri06} and at least 0.35 $\mu$s for the Behlke switch; these values apply to gaps of up  to 4 cm, or
of 16 cm, respectively. However, as these numbers show much jitter, they give only an indication for the
real timing of the voltage pulse over the gap. Therefore we have not corrected the timing $T$ in the figure
captions for the delays of the switches.

The voltage is measured with a Northstar PVM4 voltage probe. The current, in gaps of 4 cm or smaller, is
measured via a divided cathode \cite{gra87} with an inner diameter of 10 cm, details are given in
\cite{bri06}. In the 16 cm gap a different procedure needs to be followed and the stainless steel divided
cathode is replaced by a stainless steel disk of 34 cm in diameter, see figure \ref{expsetupoverview}. This
ensures that the total streamer current is collected: the distance between anode and cathode must be less
than twice the plate's diameter \cite{wetz89}. The plate is connected to the bottom of the vessel through
four parallel resistors of 33 $\Omega$ resulting in a resistance of 8.25 $\Omega$. The current is calculated
from the voltage over this resistance. To avoid sparking between the needle and the upper lid of the vacuum
vessel in the 16 cm gap, an ertalyte disk is suspended to the upper lid of the vessel (figure
\ref{expsetupoverview}). The needle is made from tungsten and has a diameter of 1 mm and tip radius of
$\sim$15 $\mu$m.

Photographs of the streamer patterns are taken with a digital 4QuikE intensified CCD camera (Stanford
Computer Optics) with a high resolution in space and a wavelength range of 200 to 800 nm. The minimal value
of the exposure time is 2 ns. The camera has a built-in delay generator to synchronize it with the event to
be recorded. With such a system, instantaneous diameters and, in particular, velocities of streamers can be
measured with high accuracy. The camera must move over a distance of 1.5 m to image the complete electrode
gaps varying from 1 to 16 cm onto the camera. For electrode gaps larger than 4 cm, the camera including the
lens can stay outside the Faraday cage and view through the ITO-window, as shown in figure
\ref{expsetupoverview}. For gaps of 4 cm or smaller, the lens has to move into the Faraday cage and the
vacuum vessel; cage and vessel are therefore adapted with tubes and windows reaching into the cage. The tube
protruding into the vacuum vessel has a BK-7 window at its end which also transmits wavelengths above 300
nm.

\subsection{Gases and pressure control}

The gases used are ambient air and nitrogen with a purity of $\approx$99.9\%, hereafter called nitrogen.
They are used in a pressure range of 0.013 to 1 bar that is measured by the Pfeiffer CMR261. The pressure
can be set on the Pfeiffer RVC300 control unit and is regulated by the Pfeiffer EVR 116 control valve. The
gases flow via Brooks Massflow controllers (5850 TR series) into  the vacuum vessel. The ambient air is the
laboratory room air of that specific moment. The atmospheric pressure is between 0.996 to 1.004 bar, the
room temperature is regulated to 21.7 $\pm$ 0.5 $^{\circ}$C and the relative humidity is 60\%. Past
experiments in a point-plane gap have shown that the effect of fluctuations of the air humidity on the
streamer is limited~\cite{veld2003}. When measuring with ambient air at atmospheric pressure, the front
viewing window is taken off to ensure sufficient ventilation. At lower pressures all vessel windows are
closed and there is no air flow into the vessel. During a measurement series no temporal changes were
observed.

The vacuum vessel is pumped down by a Speedvac ED 150 rotary vane pump to less than 0.001 bar and flushed
three times before using nitrogen. The nitrogen comes from a cylinder with a purity of 99.999\%. However,
the nitrogen is led through copper tubes into the building and via plastic tubes to the vacuum vessel.
Besides impurities picked up in the copper tubes, there are also impurities in the vessel due to components
such as the tungsten needle, the ertalyte disk, the glue and the perspex feedthrough. The vessel cannot be
baked out to remove the water in the walls. Therefore, the purity of the nitrogen is estimated to be only
$\approx$99.9\%, the rest being mainly water vapor and oxygen. A gas flow is set on the Brooks Instrument
(Readout \& Control Unit 0154) with a flow rate of 6 standard liter/min.~(slm) at 1 bar which corresponds to
a complete refreshment of the vessel in about 10 minutes. This flow rate is adjusted with pressure to keep a
refreshment time of 10 minutes, e.g., to 0.6 slm at 0.1 bar. Experiments are also performed in a
nitrogen-oxygen mixture of ratio 99.8:0.2 to compare the effects of this "impurity" level with the impurity
level of our N$_2$. The mixture is led to the vacuum vessel directly via a short plastic tube. Therefore,
the nitrogen concentration in the mixture should remain 99.8\%. Since no visible differences are found
between the mixture and our N$_2$, we conclude that the impurities of our N$_2$ are indeed about 0.1\%. The
gas composition has not been analyzed further.


\section{Measurements and results}\label{res}

\begin{figure}
\begin{center}
\vspace{0.5cm}
\begin{minipage}[b]{\linewidth}
\begin{minipage}[b]{.49\linewidth}
\includegraphics[width=7.5cm]{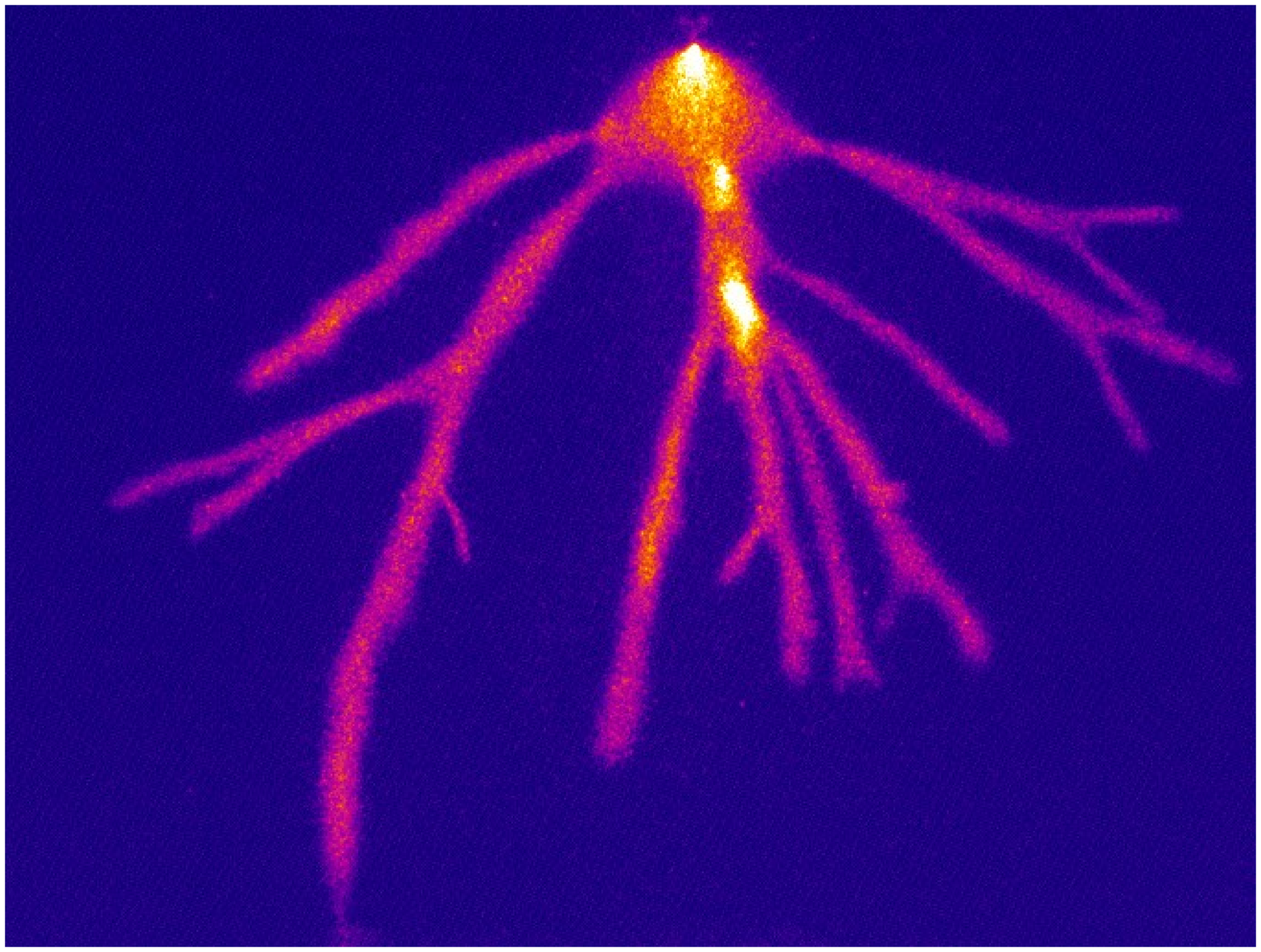}
\centering (a)
\end{minipage}\hfill
\begin{minipage}[b]{.49\linewidth}
\includegraphics[width=7.5cm]{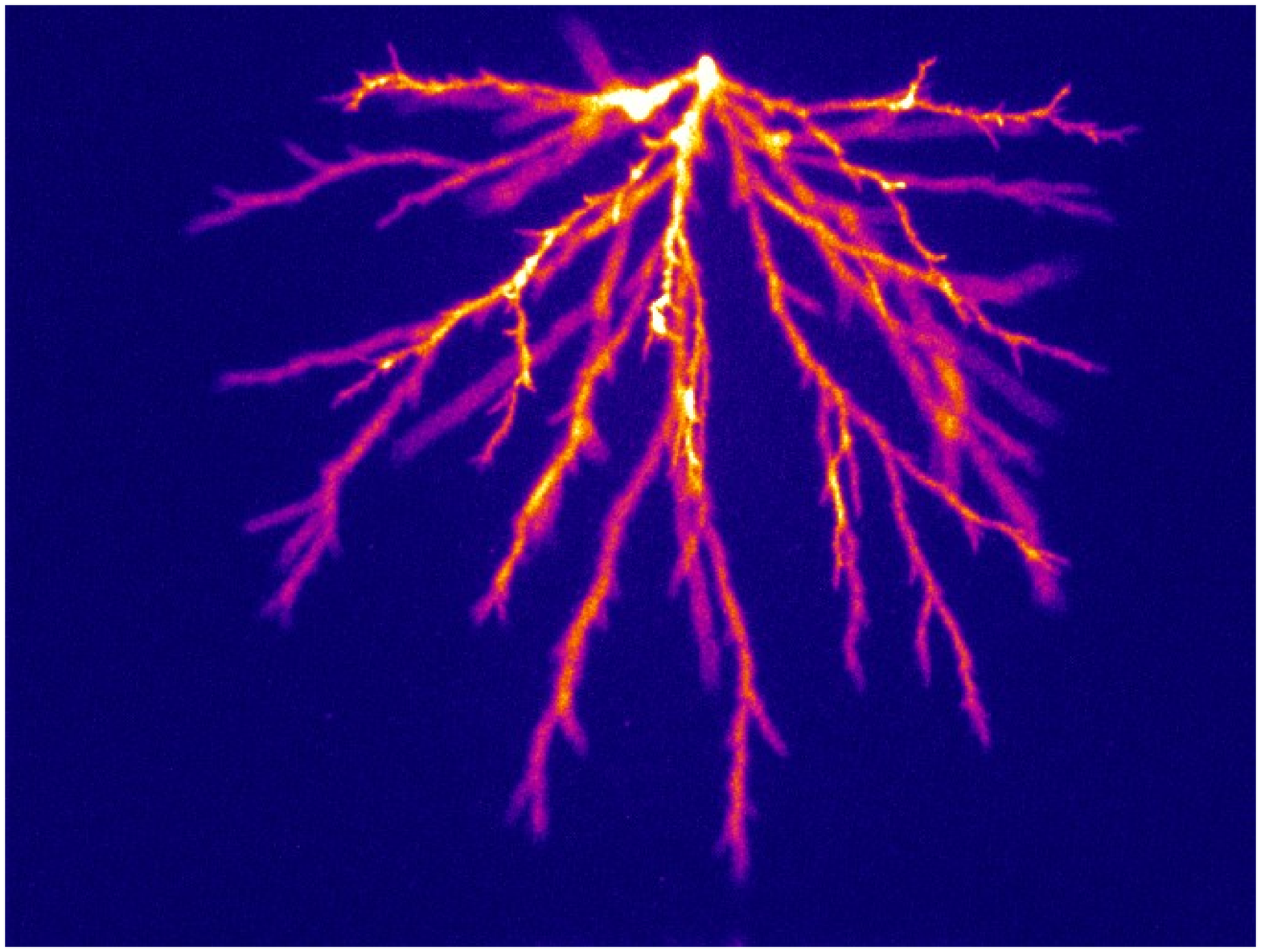}
\centering (b)
\end{minipage}\hfill
\end{minipage}
\caption{Streamers in a 4 cm gap at 0.4 bar (i.e., for $p\cdot d_{gap}=16$~mm$\cdot$bar) and at 16 kV in a)
air and b) N$_2$. In these photos, the streamers are not seen to reach the cathode because the shutter of
the camera is closed before that happens.} \label{n2airsimpel}
\end{center}
\end{figure}

A first impression of streamers in air and in our N$_2$ at 0.4 bar is given in figure \ref{n2airsimpel}. A
more detailed visual comparison of streamers in these two gases is given in \ref{app}; we also refer
to~\cite{bri08}. Briefly summarized, streamers in N$_2$ branch more frequently, but the branches extinguish
after a shorter propagation length; they are thinner and more intense \cite{bri08}. In this section we
evaluate the streamer diameter, the size of the initial ionization cloud at the electrode tip, the
propagation velocity and the streamer length between branching events; the pressure dependence and
similarity laws are discussed for all these quantities. Each subsection starts with the results in air
followed by the results in our N$_2$.

\subsection{Streamer diameter}\label{chn2airdiam}\label{chn2airschaling}

The similarity law for the streamer diameter is tested experimentally at pressures ranging from 0.013 to 1
bar. As streamer diameters and velocities depend on the electric circuit~\cite{bri06,bri08jpd,luque08jpd},
we concentrate on the minimal diameter $d_{min}$ at each pressure since this is a unique value. In general,
streamers in gaps varying from 1 to 16 cm were evaluated, while at lower pressures, the shorter gaps are too
short to form a streamer. In particular, for 0.013 bar, only a gap of 16 cm is long enough for a streamer to
form, as is illustrated in fig.~\ref{gevoeltijdsopg}.

\begin{figure}
\begin{center}
\vspace{0.5cm}
\begin{minipage}[b]{\linewidth}
\begin{minipage}[b]{.48\linewidth}
\includegraphics[height=6cm]{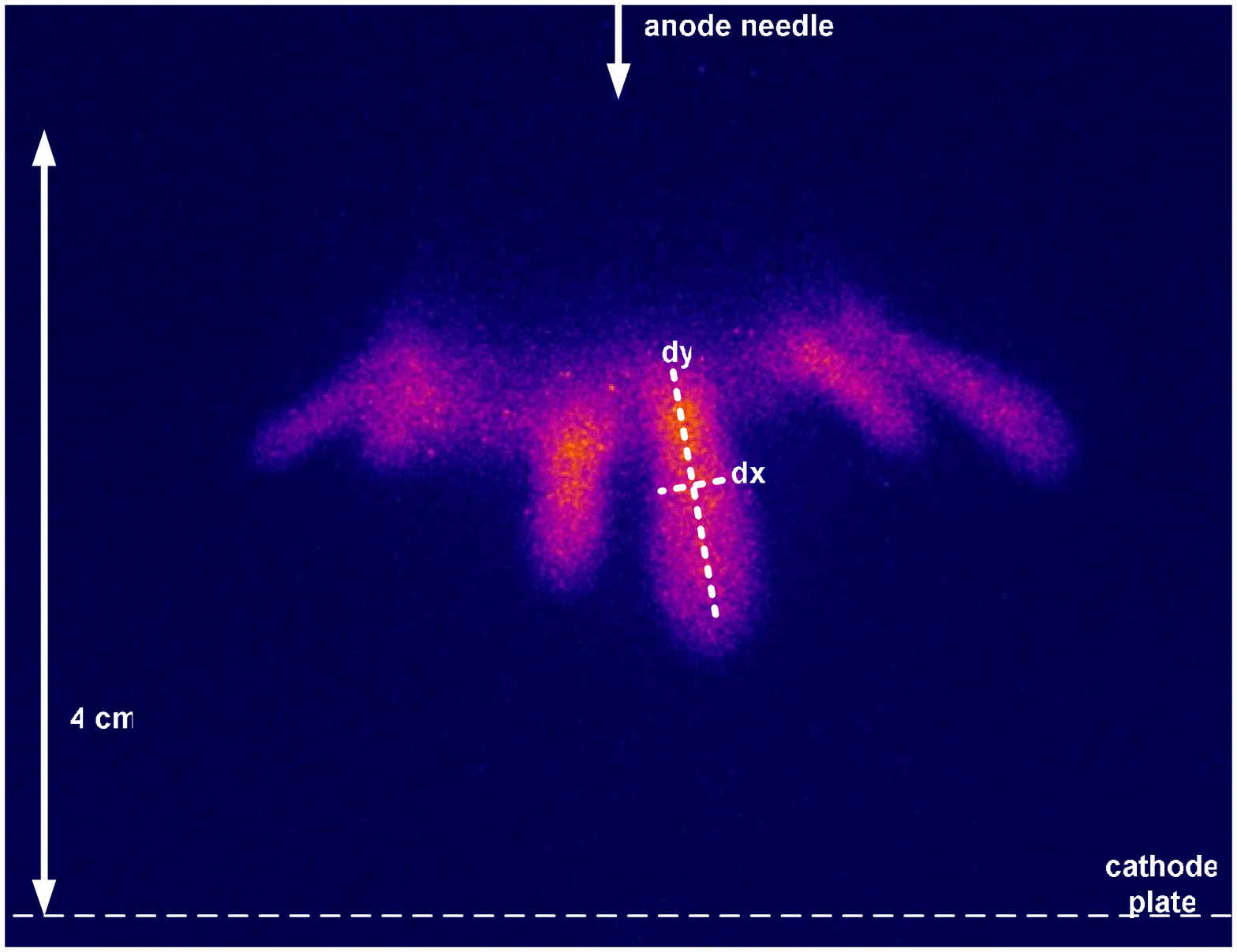}
\centering (a)
\end{minipage}\hfill
\begin{minipage}[b]{.48\linewidth}
\includegraphics[height=6cm]{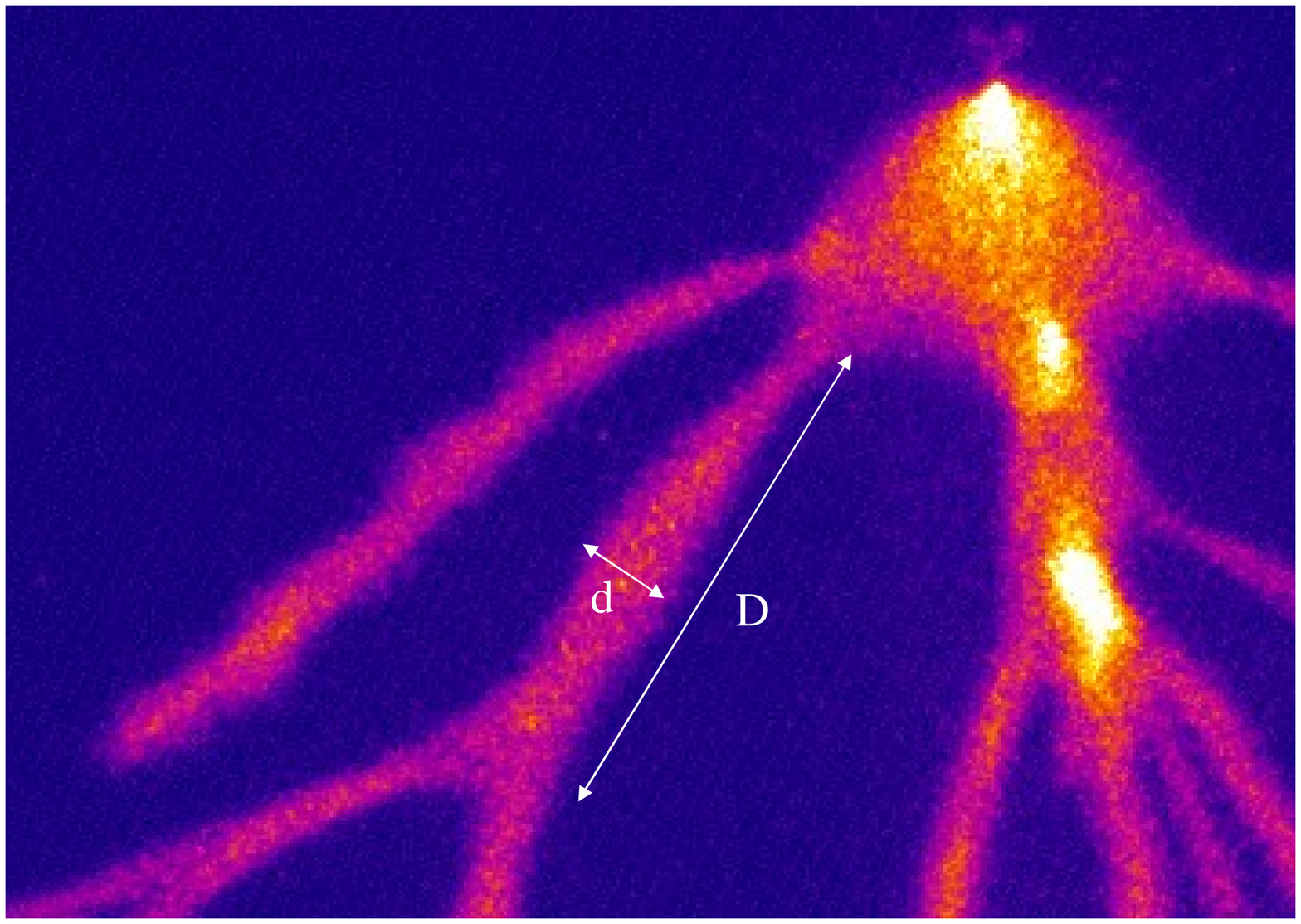}
\centering (b)
\end{minipage}\hfill
\end{minipage}
\caption{a) Positive streamers in a 4 cm point-plane gap in air at 0.1 bar and 8 kV ($p\cdot
d_{gap}=4$~mm$\cdot$bar). The delay after the pulse of the function generator is $T=1.6~\mu$s and the
exposure time is $\Delta t= 50$ ns. The full width $dx$ at half maximum determines the diameter, and the
ration of $dy$ over $\Delta t$ the velocity. b) The distance $D$ between branching events and the streamer
diameter $d$. This picture is a zoom into the upper left corner of Fig.~\ref{n2airsimpel}a.} \label{figdiag}
\end{center}
\end{figure}

The streamer diameters are determined from iCCD-photographs such as shown in Fig.~\ref{figdiag}a. The
diameter is obtained as the full width at half maximum of the cross-section d$x$, averaged over a number of
lines in the d$y$ direction for as long as the streamer channel is straight; by averaging over a number of
cross sections, stochastic fluctuations of the number of counted photons per pixel are averaged
out\footnote{This averaging procedure was not yet followed in our older proceedings
article~\cite{tanjaicpig}, and the diameter data presented there is much less extensive and less accurate
than the one presented here.}. Note that the direct visual inspection of the color-coded data representation
in the figures can occasionally be misleading; it can mimic different diameters than the evaluation of the
original data. Care is taken that figures of single, in-focus streamers at a place without return stroke or
electrode effects are evaluated. Therefore, the camera's exposure time is adjusted in such a way that only
the primary streamer during its flight is photographed. Diameters in a 16 cm gap are determined from
photographs where the camera has been moved physically towards the discharge; an example of such a close-up
is shown in panel (d) of Fig.~\ref{n2airvergelijk}. As discussed in depth in~\cite{bri06}, the diameter must
contain a sufficient number of pixels to avoid artifacts in the evaluation; in the present measurements, the
diameter covers at least seven pixels.

\begin{figure}
\begin{center}
\vspace{0.5cm}
\includegraphics[width=12cm]{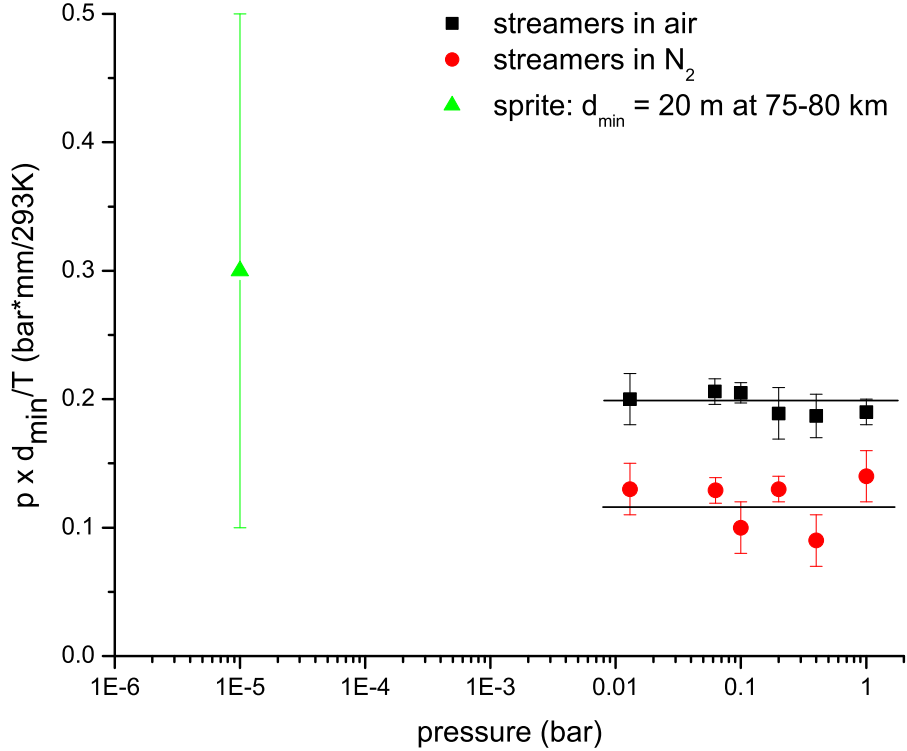}
\caption{Reduced minimal streamer diameter $d_{min}\cdot p/T$ as a function of pressure $p$; here $T$ is
temperature and 293 K is room temperature. Black squares: experimental results in air. Red dots:
Experimental results in our N$_2$. Green triangle: minimal sprite diameter at 80 km height
from~\cite{gerken}.} \label{n2airEddie}
\end{center}
\end{figure}

The similarity law for the streamer diameters is tested in figure \ref{n2airEddie}; here the minimal
streamer diameters $d_{min}$ at each pressure are multiplied by the pressure $p$, and the reduced diameters
$p\cdot d_{\rm min}$ are plotted as a function of $p$; black squares indicate air and red dots our N$_2$.
The error bars indicate the standard deviation of about 10 different measurements of streamer diameters in
each combination of pressure and gap size as described above. Systematic errors are estimated to be less
than the stochastic errors due to noise, etc. A linear fit with the constant value $p\cdot d_{\rm min}$ =
0.20 $\pm$ 0.02 mm$\cdot$bar matches the data points very well. Hence the experiments show, that the minimal
streamer diameters scale very well with pressure.

The green triangle in figure \ref{n2airEddie} presents an estimate of the minimal diameter of a sprite
discharge at an altitude of 75-80 km \cite{gerken}. Several values for the transverse extension of the
sprite channels are mentioned in this paper, and we take the smallest value of 20 $\pm$ 13 m that is
reported at a height of 80 km. We use this value as an estimate and upper bound of the minimal diameter. At
80 km height, the pressure is about 10$^{-5}$ bar and the temperature is about -83$^\circ$C \cite{binas}.
Here one needs to account for the fact that the physically determining factor is not pressure and $p\cdot
d$, but density and $n\cdot d = p\cdot d/(k_B T)$, as discussed in section~\ref{Townsend}. Therefore, in
figure \ref{n2airEddie}, $p\cdot d/T$ is plotted. When the sprite diameter of $20\pm13$~m is extrapolated to
room temperature, we find $p\cdot d/T= 0.3\pm 0.2$ mm$\cdot$bar/(293 K). This reduced diameter is in the
same range as the minimal streamer diameter of 0.20 $\pm$ 0.02 mm$\cdot$bar in air in our laboratory
experiments. (A referee suggests to rather use the data of NASA's MSIS~\cite{MSIS} with a pressure of
$1.18\cdot10^{-5}$~bar and a temperature of -105$^\circ$C at 80 km height. This leads to a somewhat larger
reduced diameter of $p\cdot d/T= 0.4\pm 0.25$ mm$\cdot$bar/(293 K).) The sprite diameter can be uncertain
due to instrumental broadening of the telescope and to uncertainties about the actual sprite height and
local air density. Furthermore, sprites propagate through different density regimes, and the reported sprite
diameters might not be minimal. Therefore, we look forward to further sprite data for comparison.

The similarity law is also tested for minimal streamers in N$_2$ (99.9\%). Here we fit the data with the
constant $p\cdot d_{\rm min}$ = 0.12 $\pm$ 0.03 mm$\cdot$bar. The statistical fluctuations in N$_2$ in
figure \ref{n2airEddie} are larger and the error bars at several pressures do not intersect with the fitted
line. This effect can already be seen by eye on the photos: the streamers in nitrogen at 0.4 bar in
Fig.~\ref{n2airsimpel}~(b) are not a factor 2.5 thicker than the ones at 1 bar in
Fig.~\ref{n2airuiteinde}~(b). This effect could be due to fluctuating impurity concentrations in our N$_2$.

\subsection{Diameter of the initial ionization cloud at the electrode tip}

\begin{figure}
\begin{center}
\begin{minipage}[b]{\linewidth}
\begin{minipage}[b]{.33\linewidth}
\includegraphics[width=5cm]{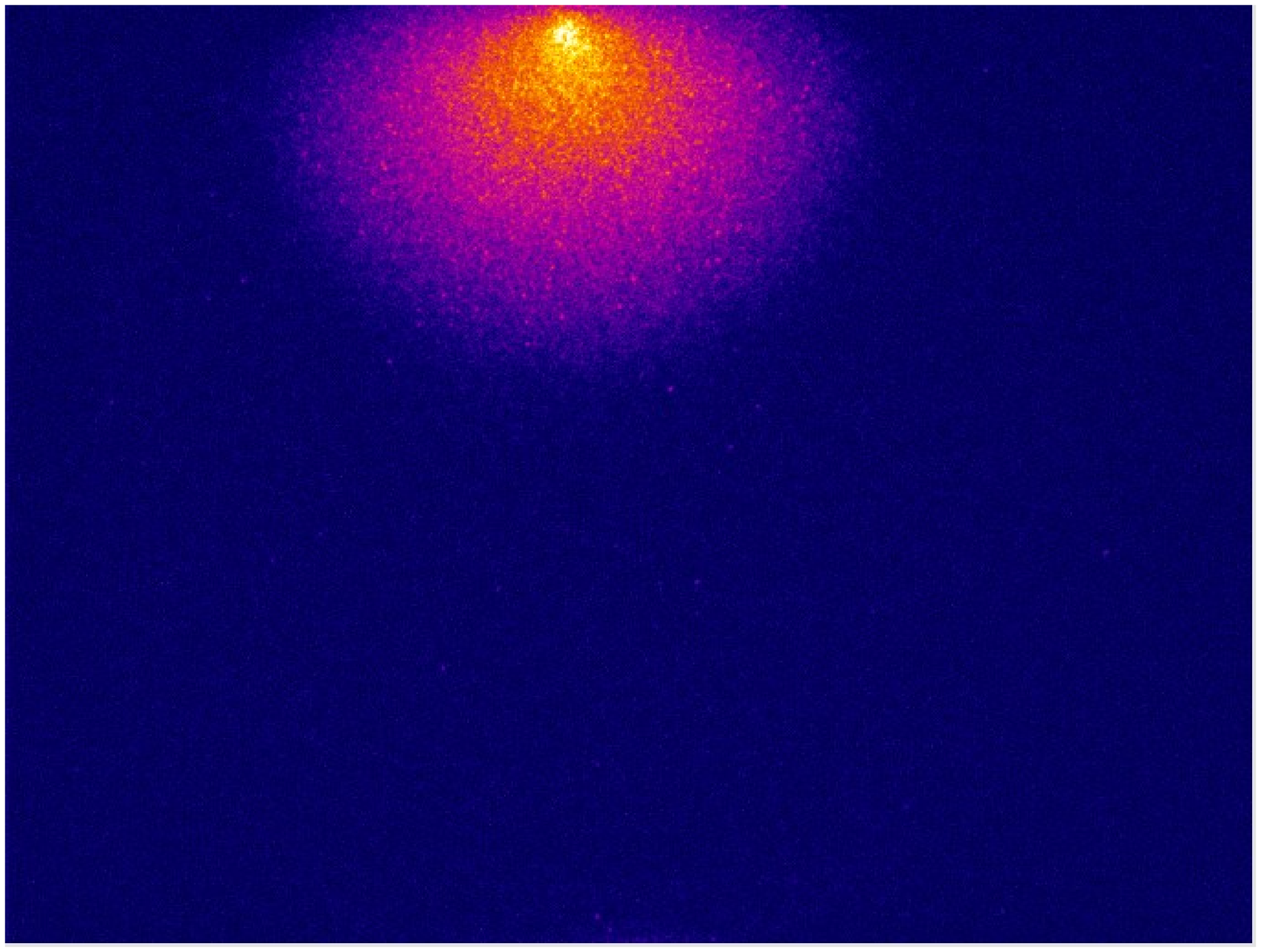}
(a): $T=1.4 \mu$s, $\Delta t =0.2\mu$s
\end{minipage}\hfill
\begin{minipage}[b]{.33\linewidth}
\includegraphics[width=5cm]{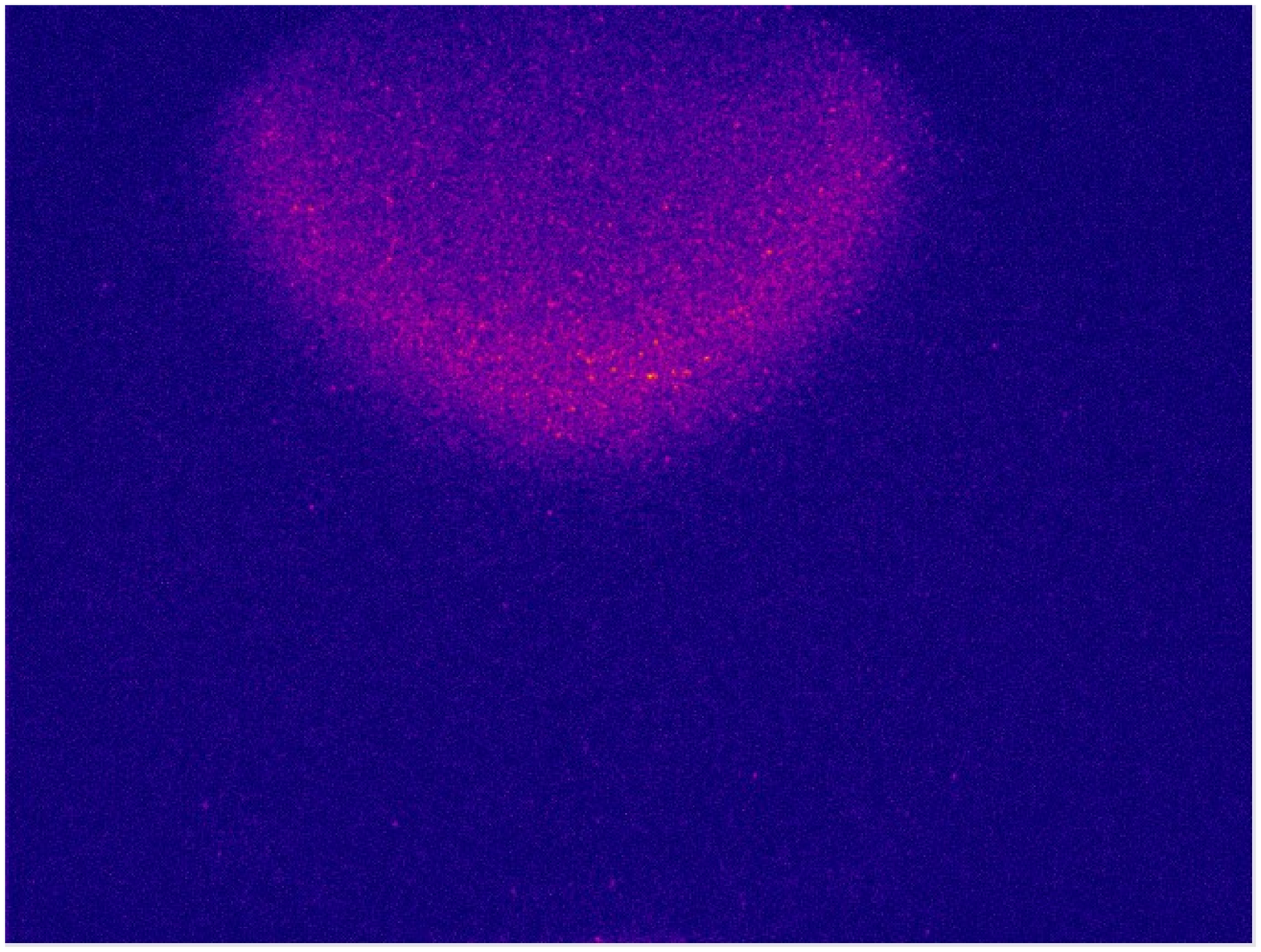}
(b): $T=1.6 \mu$s, $\Delta t =0.2\mu$s
\end{minipage}\hfill
\begin{minipage}[b]{.33\linewidth}
\includegraphics[width=5cm]{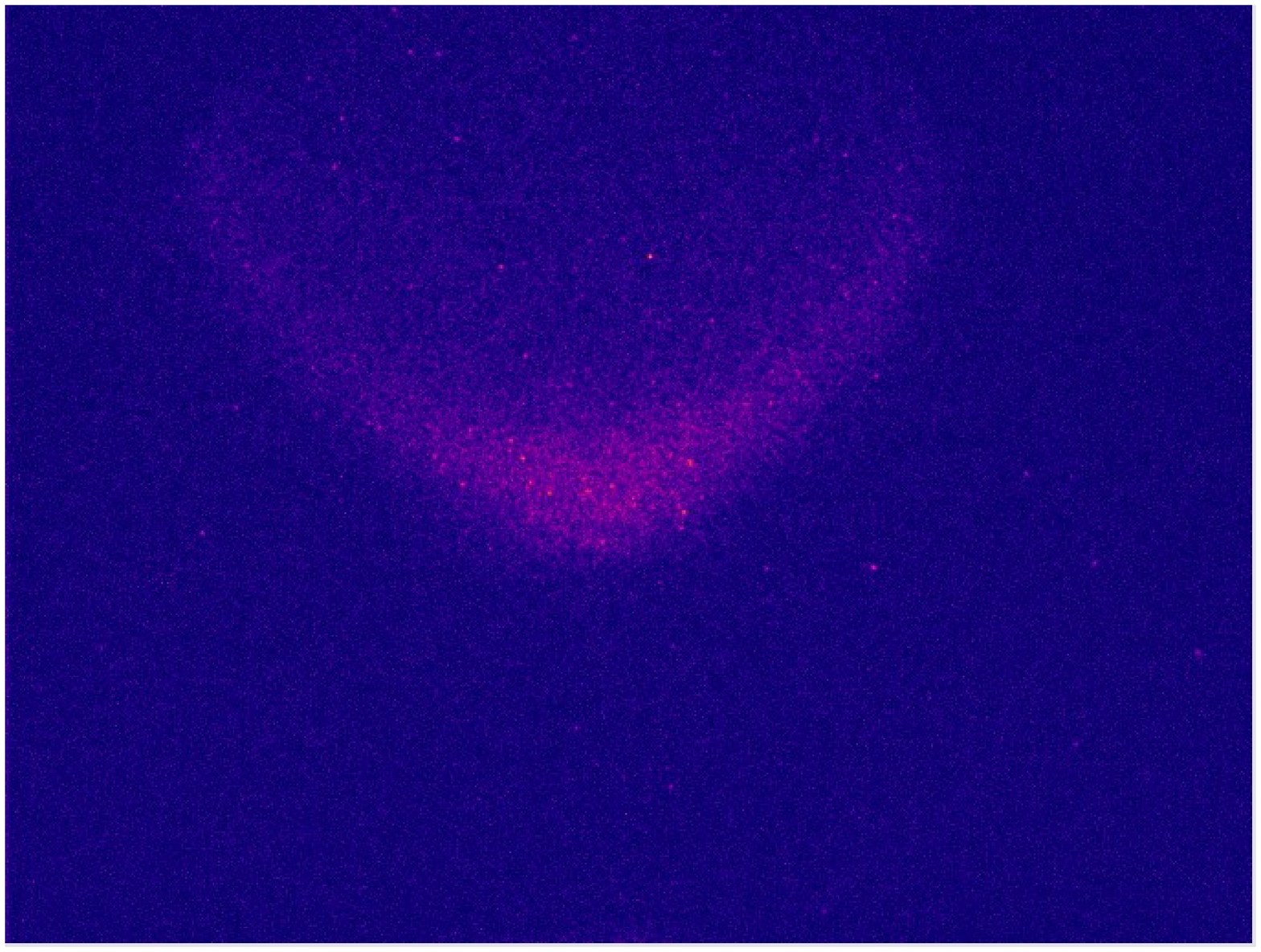}
(c): $T=1.8 \mu$s, $\Delta t =0.2\mu$s
\end{minipage}\hfill
\end{minipage}\vspace{2ex}
\begin{minipage}[b]{\linewidth}
\begin{minipage}[b]{.33\linewidth}
\includegraphics[width=5cm]{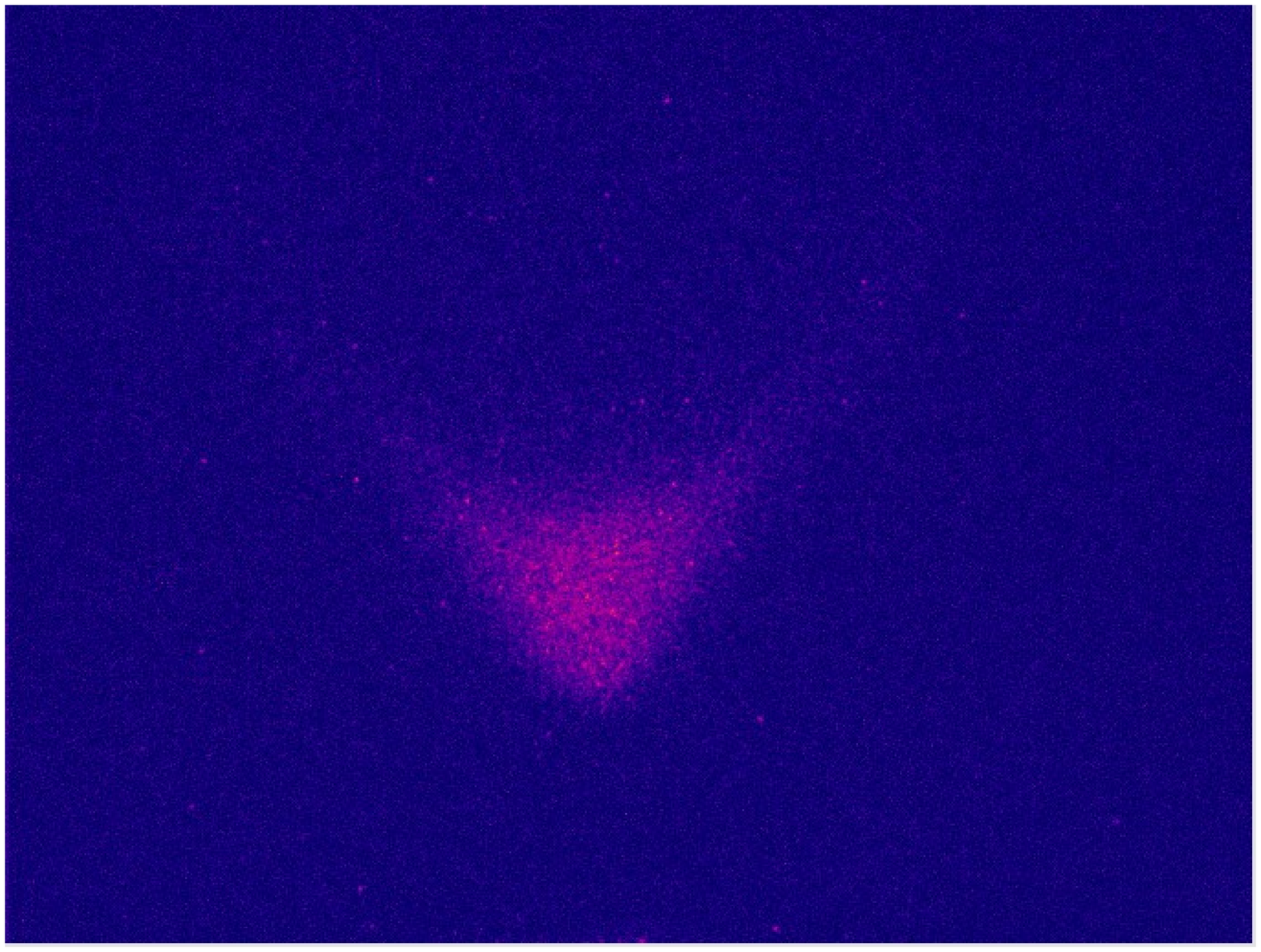}
(d): $T=2.0 \mu$s, $\Delta t =0.2 \mu$s
\end{minipage}\hfill
\begin{minipage}[b]{.33\linewidth}
\includegraphics[width=5cm]{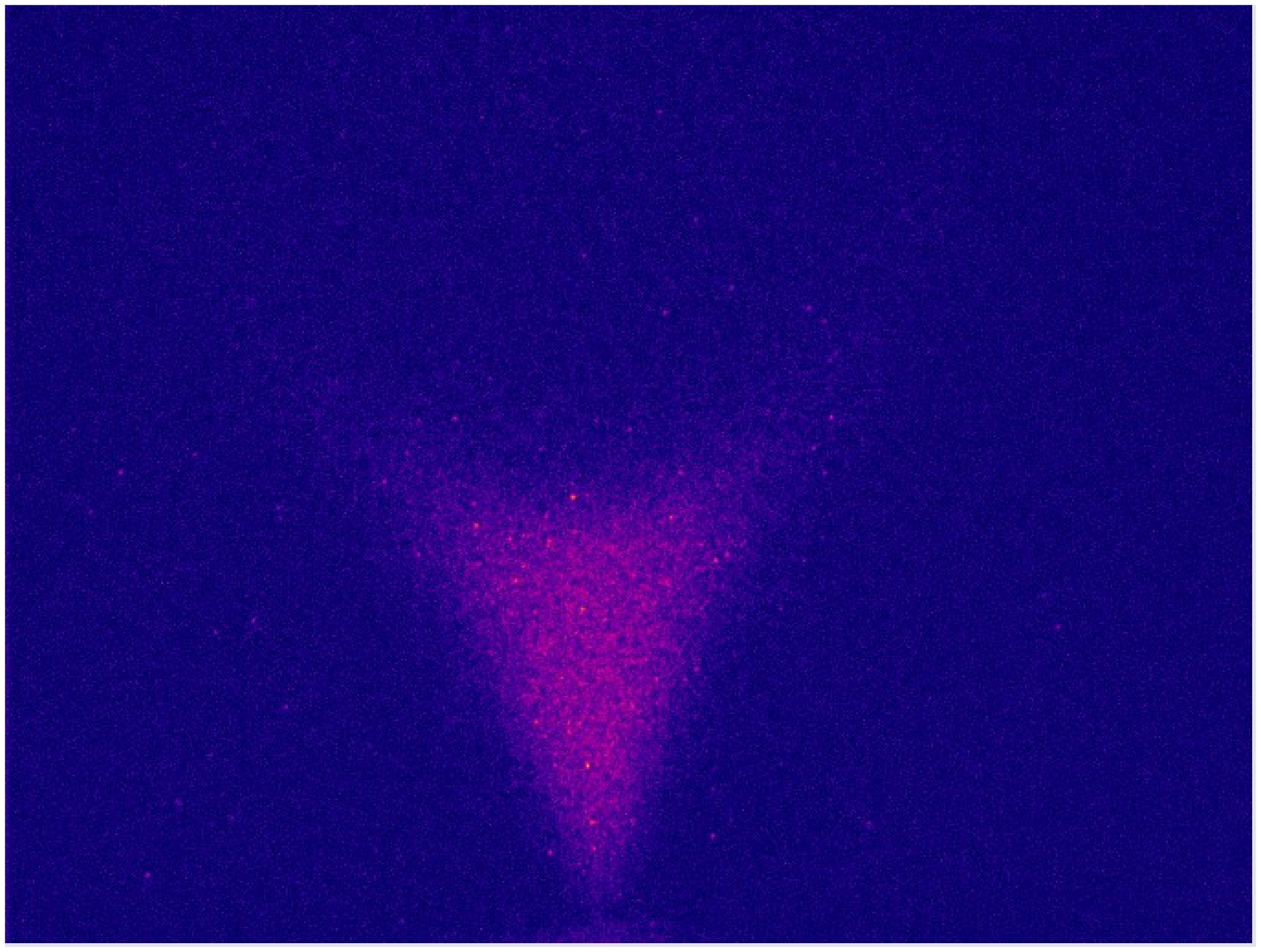}
(e): $T= 1.8 \mu$s, $\Delta t =0.3\mu$s
\end{minipage}\hfill
\begin{minipage}[b]{.33\linewidth}
\includegraphics[width=5cm]{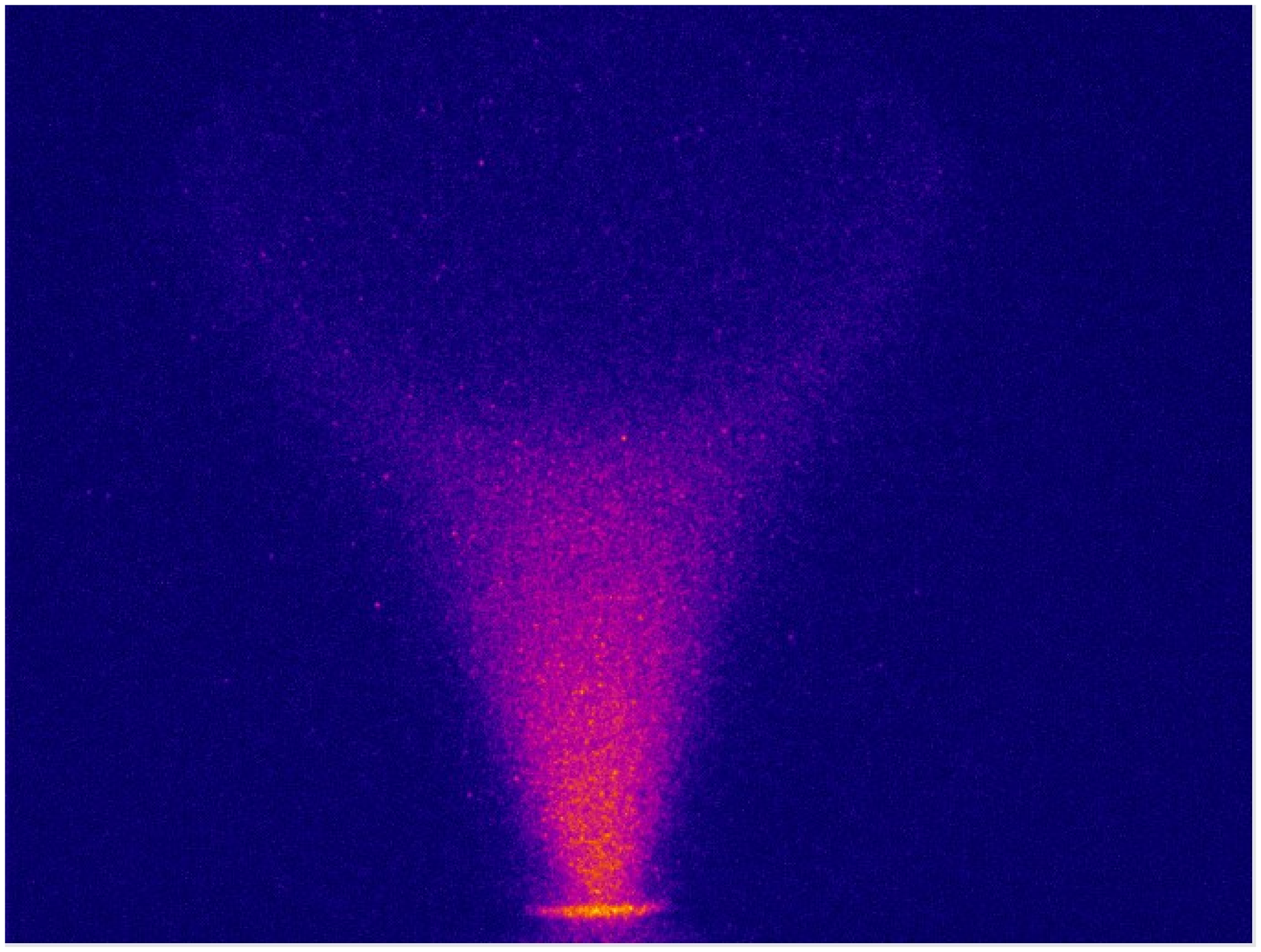}
(f): $T=1.7 \mu$s, $\Delta t =0.4\mu$s
\end{minipage}\hfill
\end{minipage}
\caption{Time resolved photographs of streamer evolution in a 16 cm gap at 0.015 bar ($p\cdot
d_{gap}=2.4$~mm$\cdot$bar) in a nitrogen-oxygen mixture of ratio 99.8:0.2 at a voltage of 5 kV. The camera's
opening time $T$ and exposure time $\Delta t$ are indicated in each figure. Since the actual streamer
inception time relative to $T$ is subject to some jitter between different experiments, the photographs are
ordered in a plausible temporal sequence. The panels show (a) a light emitting inception cloud at the
electrode tip that (b,c) evolves into a shell from which (d) one streamer emerges that (e) propagates
towards and (f) reaches the plate electrode. A similar set of figures in a 4 cm gap in air at pressures of
0.1 and 0.4 bar is presented in~\cite{bri08_2}.
} \label{gevoeltijdsopg}
\end{center}
\end{figure}

The time resolved streamer measurements at low pressure also allow us to briefly describe its early stages
of evolution. The discharge starts with the formation of a glowing ionization cloud at the tip of the anode
needle. The further evolution in nitrogen in a small reduced gap length of $p\cdot L=2.4~{\rm mm \cdot bar}$
is shown in figure~\ref{gevoeltijdsopg} where $L=16$~cm is the gap length; a similar evolution in air is
presented in \cite{bri08_2}. (For comparison, figure~\ref{vel} has $p\cdot L=160~{\rm mm \cdot bar}$.) The
heights and widths of these ionization clouds are given in table \ref{tabn2aircloud}; they are measured from
a snapshot where the cloud is highest and widest and at the position where the intensity has decreased to
20\% of its maximal value. The dimensions are mainly evaluated before streamers emerge. Due to the large
time jitter in the cloud initiation, it is a matter of trial-and-error to obtain a good photograph, and it
is very time consuming to create a series of photos as in \ref{gevoeltijdsopg}. As a consequence, table
\ref{tabn2aircloud} is based on a quite limited data set. At pressures above 0.2 bar, the clouds can not be
resolved accurately from the available photos. (A lens system to zoom in on these details was not available
during the measurements.)

\begin{table}[t]
\begin{center}
\centering
\begin{tabular}{|c|c|c|c|c|c|c|}\hline
$p$ (bar) & gap (cm) & $U$ (kV)    & height $h$ (mm)      & width $w$ (mm) & $h\cdot p$ (mm$\cdot$bar) & $h/w$\\
\hline \hline
0.013     & 16         & 2             &  30     & 45 & 0.4& 0.7\\ 
        &           & 3           & 45    &  73 & 0.6 & 0.6 \\
        &           & 4           & 38-50    &  75-90 & 0.5-0.7 & 0.5-0.6\\
\hline
0.1      & 4    & 2            &  5         &    3 & 0.5 & 1.7\\
        &           & 5          &  4-9       &  5-10 & 0.5-1  & 0.9 \\
        &          & 8              &  8-9   & 10-16 & 0.8-0.9 & 0.6-0.8\\
\hline
0.2       & 4      & 5                & 3       & 2 & 0.6 & 1.5 \\
        &          & 10              & 5-8       &  9-15 & 1-1.6 & 0.5-0.6\\
        &          & 20              & 20         &  24-28 & 4 & 0.8\\
\hline
\end{tabular}
\caption{Height $h$ and width $w$ of the inception cloud at the electrode tip in air. The error bar of these
data is up to 50\%.}\label{tabn2aircloud}
\end{center}
\end{table}

The dimensions of the cloud in air increase with decreasing pressure or increasing voltage. The cloud is
generally oblate with an aspect ratio of $h/w \approx$ 0.7. At 0.1 and 0.2 bar, a different aspect ratio of
$h/w \approx$ 1.6 is seen at the lowest voltages. The height of the cloud at the inception voltage roughly
scales with inverse pressure; this leads to a reduced value of $h\cdot p = 0.5 \pm 0.2$~mm$\cdot$bar.

In N$_2$, the cloud is smaller and less regular than in air, as can, e.g., be seen in figure
\ref{n2airsimpel}. Only at the lowest pressures of 0.015 bar in a 16 cm gap, the sphere is clearly observed
with a size similar to the one in air, cf.~Fig.~\ref{gevoeltijdsopg}.

\subsection{Streamer velocities}

\begin{figure}
\begin{center}
\vspace{0.5cm}
\begin{minipage}[b]{\linewidth}
\begin{minipage}[b]{.33\linewidth}
\includegraphics[width=5cm]{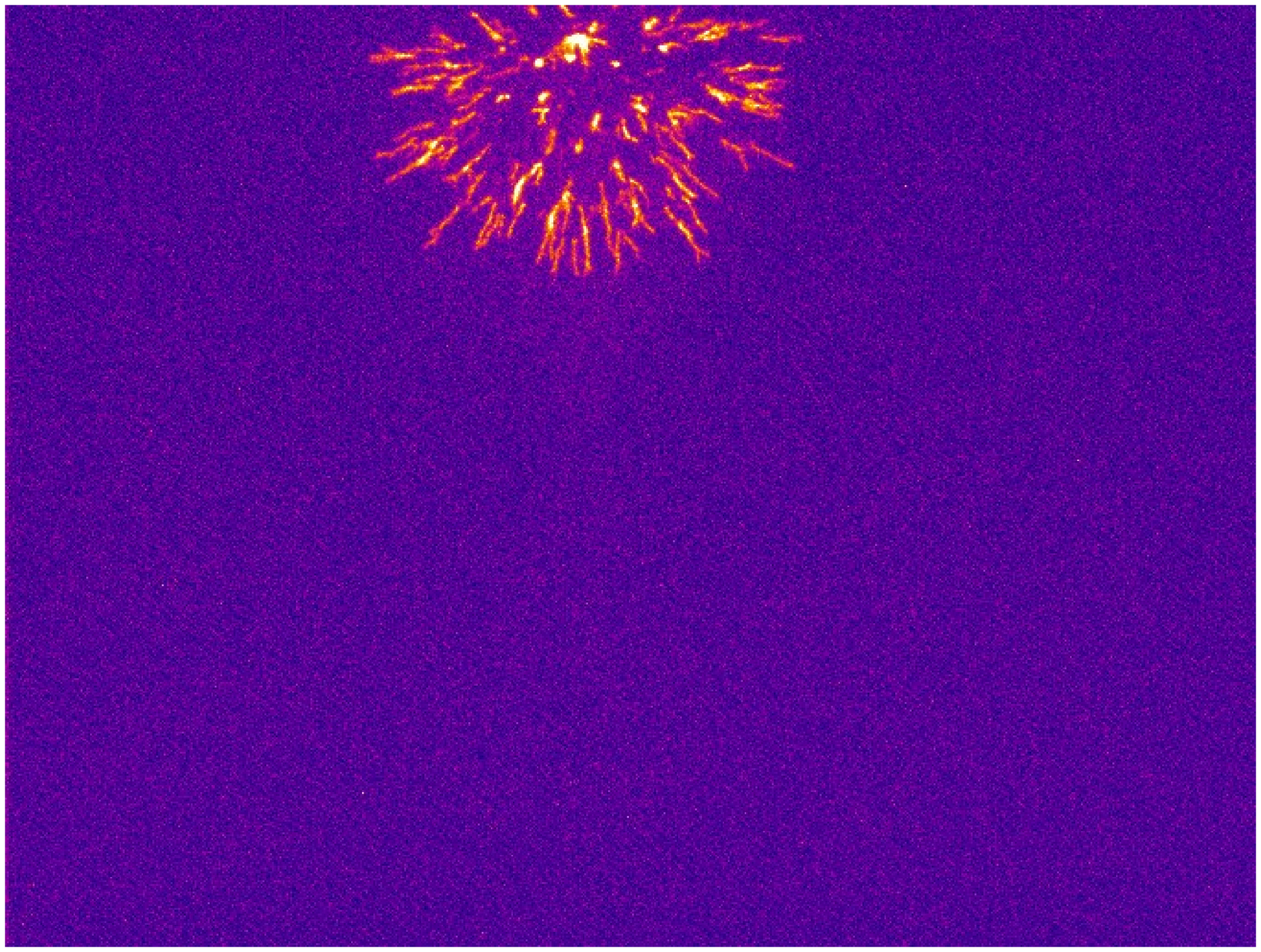}
\centering (a) $T = 2 \mu$s, $\Delta t = 0.1 \mu$s
\end{minipage}\hfill
\begin{minipage}[b]{.33\linewidth}
\includegraphics[width=5cm]{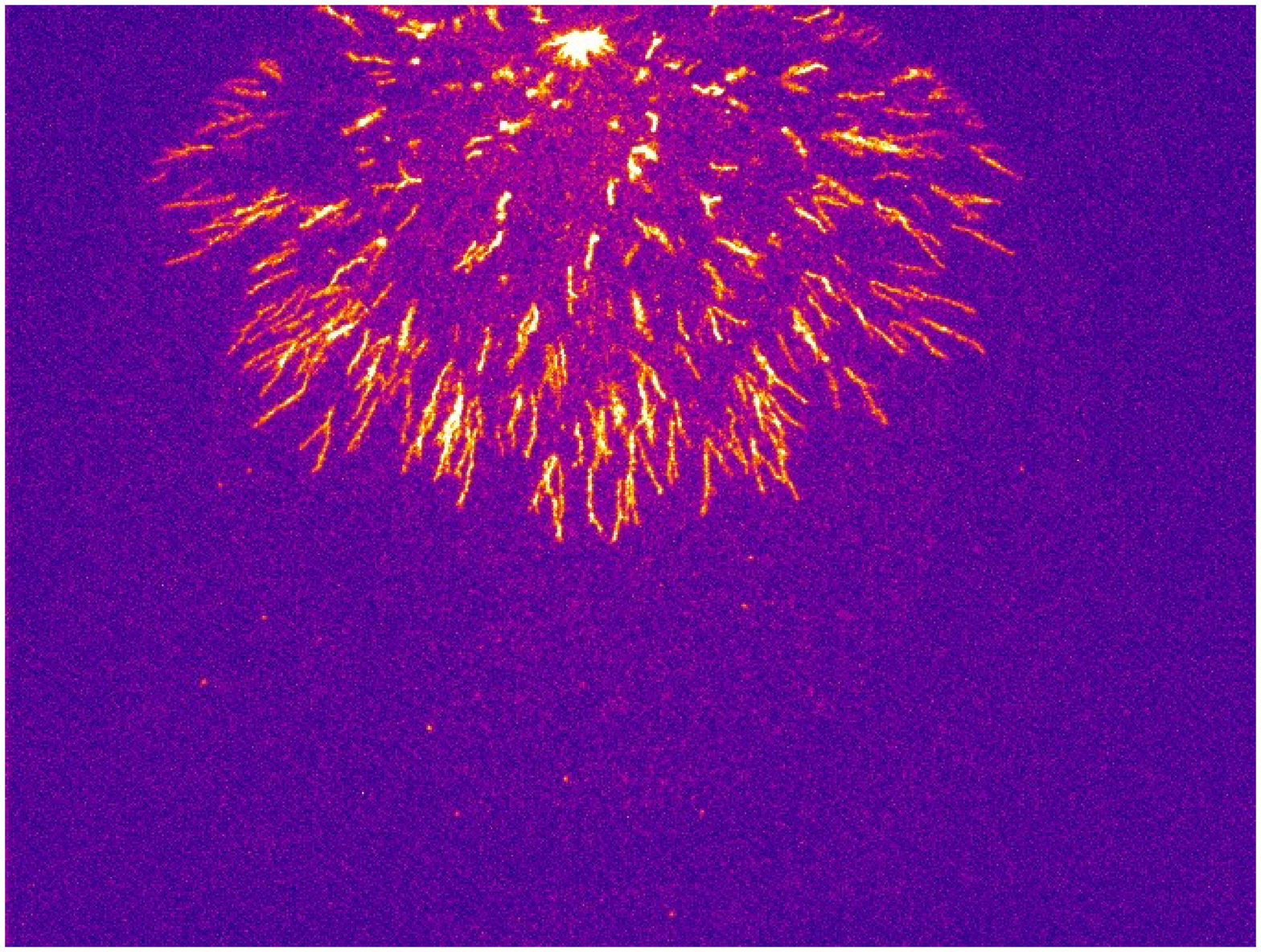}
\centering (b) $T = 3 \mu$s, $\Delta t = 0.5 \mu$s
\end{minipage}\hfill
\begin{minipage}[b]{.33\linewidth}
\includegraphics[width=5cm]{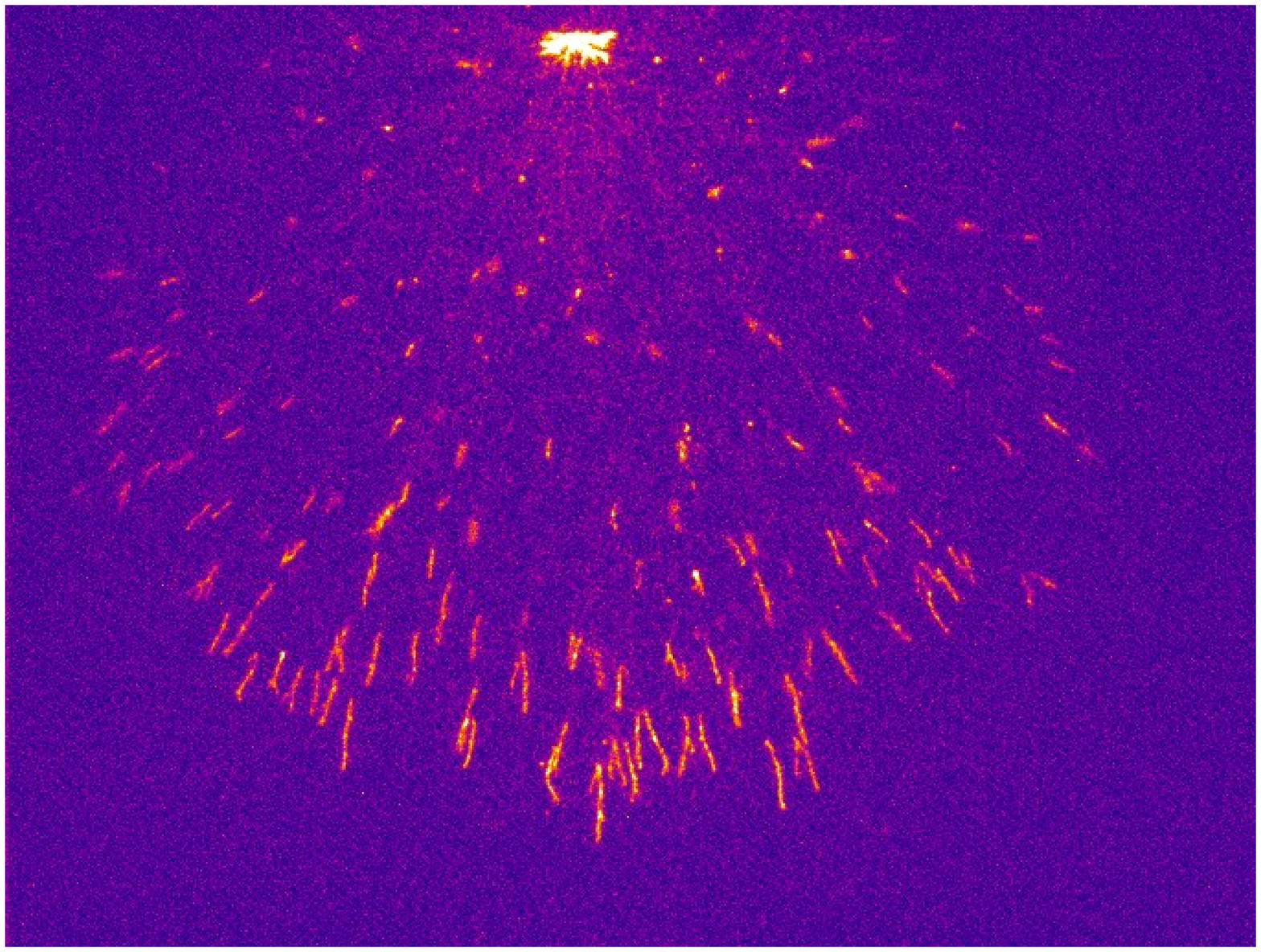}
\centering (c) $T = 5 \mu$s, $\Delta t = 0.5 \mu$s
\end{minipage}\hfill
\end{minipage}
\caption{Time resolved photographs of streamers in a 16 cm gap in our N$_2$ at 1 bar at a voltage of 49 kV
($p\cdot d_{gap}=160$~mm$\cdot$bar). The gate delay $T$ and exposure time $\Delta t$ are indicated in each
panel. The streamers reach the plate electrode some time after the exposure of panel~(c). When the same
experiment is performed in air at 1 bar, the streamers do not cross the gap.}\label{vel}
\end{center}
\end{figure}

Streamer velocities are evaluated from single photographs like panel (a) of figure~\ref{figdiag}. The method
is based on the observation that for exposure times as short as 1 ns, the actively growing streamer head in
air and nitrogen is visible as a glowing ball~\cite{ebe06}, this feature recently also has been demonstrated
for sprites~\cite{McHarg}. Extended lines therefore indicate the trace of the propagating streamer head
within a longer exposure time. Delay $T$ and exposure time $\Delta t$ of the photographs are chosen in such
a way that velocities conveniently can be determined, see below. We evaluate the velocities here only in the
16 cm gap and at three different locations; near the anode tip, halfway the gap and near the cathode plate.
The velocities are determined from the longest streamers in each picture since they propagate most parallel
to the camera's focal plane. (This constraint could be dropped in future measurements when using
stereographic imaging~\cite{Nijdam08}.) Note that the velocities are related to the
diameters~\cite{bri06,bri08jpd,luque08jpd}. However, the diameters cannot be accurately determined from the
full photographs of the 16 cm gap as the camera resolution is insufficient, cf.\ the extensive discussion of
imaging artifacts in~\cite{bri06} that is briefly recalled in section~\ref{chn2airdiam}.

Typical exposure times $\Delta t$ for the velocity measurements are 50 to 500 ns. These values are chosen in
such a way that the propagation distance is short enough to contain little or no branching, but also long
enough to limit relative errors in d$y$ and $\Delta t$ due to the extension of the streamer head, due to the
emission time of the excited C-state of nitrogen ($\sim$2 ns) that emits the light, and due to inaccuracies
of the camera's gate time $\Delta t$. Therefore the exposure time is adjusted such that the streamer trace
d$y$ is about 1 to 4 cm in the 16 cm gap as in Fig.~\ref{vel}. At 0.015 mbar the development of the streamer
is so limited that its velocity cannot be obtained, cf.~figure~\ref{gevoeltijdsopg}; a much larger vessel
would be required for this purpose.

\begin{figure}
\begin{center}
\vspace{0.5cm}
\begin{minipage}[b]{\linewidth}
\begin{minipage}[b]{.48\linewidth}
\includegraphics[height=6cm]{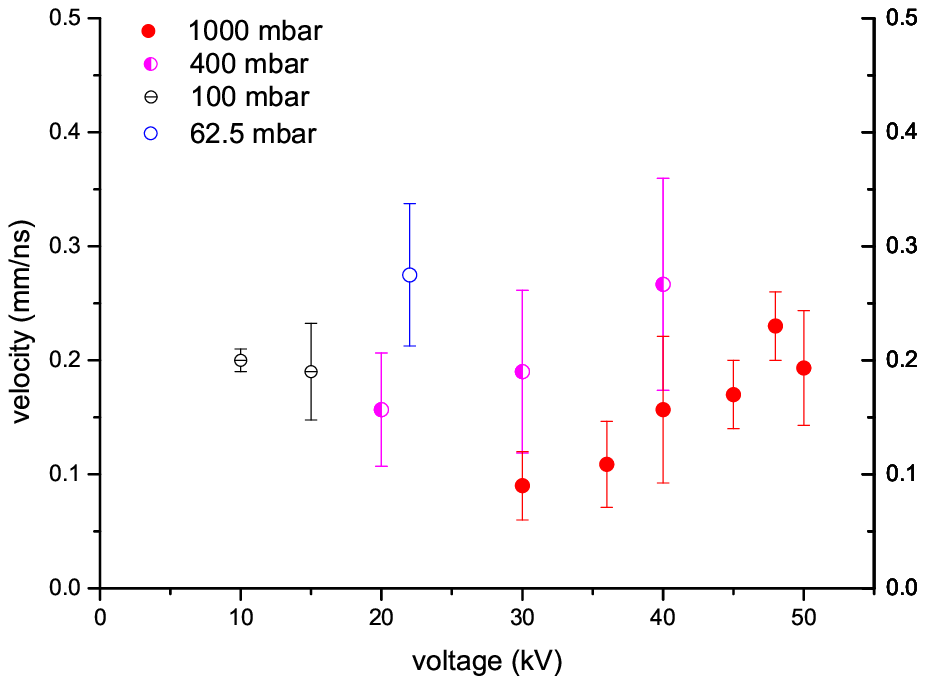}
\centering (a)
\end{minipage}\hfill
\begin{minipage}[b]{.48\linewidth}
\includegraphics[height=6cm]{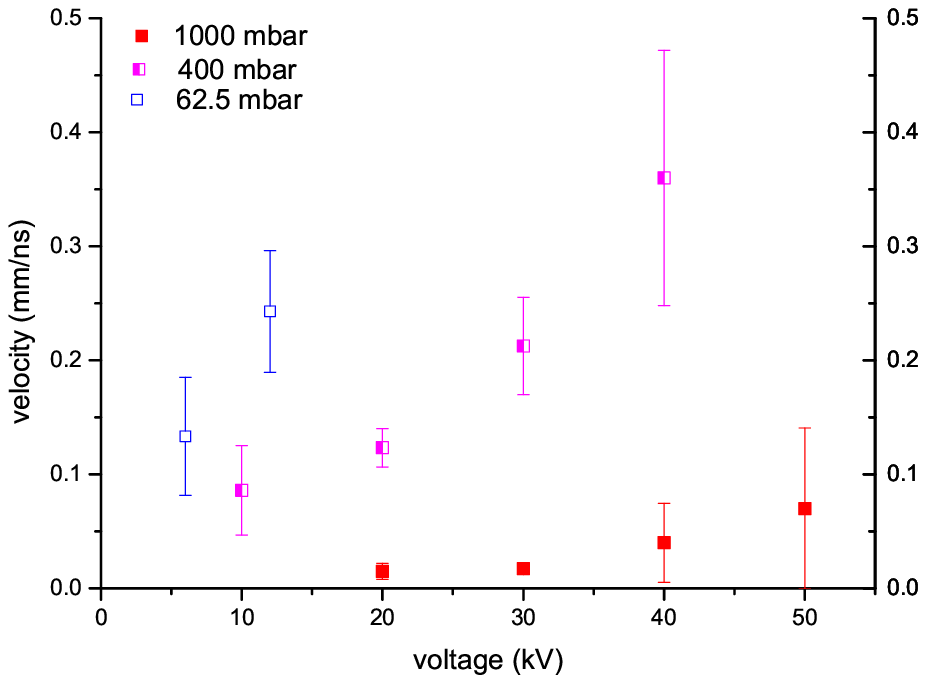}
\centering (b)
\end{minipage}\hfill
\end{minipage}
\caption{The propagation velocity of streamers in a 16 cm gap at different pressures a) in air and b) in our
N$_2$ (1 mm/ns = $10^6$ m/s). The bars indicate the velocity range observed at different locations in the
gap.} \label{n2airvel}
\end{center}
\end{figure}

The comparison of panels (a) and (b) in figure~\ref{vel} shows, that the propagation length is roughly the
same while the exposure time is five times longer. This demonstrates the general feature that the velocity
decreases when the streamer propagates down the gap into regions with lower background field; in
Fig.~\ref{vel} the velocity change between different gap locations is in fact larger than in other cases.
Fig.~\ref{n2airvel} presents the range of streamer velocities as a function of pressure and voltage. The
bars in the figure extend from the lowest velocity (typically measured near the cathode plate) to the
highest velocity (typically near the anode tip).

The jitter in the discharge inception makes it here difficult as well to obtain images that can be analyzed
with confidence. Therefore the number of time resolved photographs per parameter set is limited to only a
few. Furthermore, at low voltages, one must be sure that the streamer on the image finishes because the
camera closes, and not because the streamer stops anyway. In air this constraint is particularly severe, as
the maximal streamer length is typically shorter than in nitrogen, as is shown in
figure~\ref{n2airvergelijk}. For all these reasons, it is uncertain whether the velocity of streamers of
minimal diameter is contained in our measured data set. In air at 1 bar the minimal velocity is of the order
of 0.1 mm/ns = $10^5$ m/s; far from the needle electrode it even decreases to $6\cdot10^4$~m/s. A value of
$10^5$~m/s is also found as the minimal velocity in a 40 mm gap \cite{bri08jpd}. At 0.4 bar the smallest
velocity takes the higher value of $1.5\cdot10^5$~m/s, but then the diameter is clearly not minimal, but has
the value of $p\cdot d = 0.6$~mm$\cdot$bar, cf.~Fig. 6.9 a in \cite{briPhD}.

In our N$_2$ the situation is more complicated. The velocities at 1 bar are much lower than in air, e.g.,
from figure~\ref{vel}~(c), one can easily extract a velocity of $3\cdot 10^4$~m/s. But at lower pressure the
velocities in nitrogen are the same or even higher than in air; here a pressure dependent impurity level
might play a role.

\subsection{Branching length}\label{chn2enairdistance}

Streamer branching can be characterized by the branching angle and the streamer length between branching.
For the determination of the branching angle through stereographic imaging, we refer to~\cite{Nijdam08}.
Here we measure the length $D$ of the streamer from one branching event to the next. We relate this length
to the streamer diameter $d$ in the branching ratio $D/d$, both quantities are illustrated in panel (b) of
figure \ref{figdiag}. For this evaluation, photographs must show the complete discharge as well as accurate
diameters without instrumental broadening, therefore only data from a 4 cm gap are evaluated. The diameter
is measured just before the branching of the channel; the streamer length between branching events is
measured as the starting point of one streamer branch to the starting point of the next one. Evaluations in
the region close to the electrode tip are mostly avoided since the tip region is frequently overexposed and
therefore the actual starting point of the streamer can not be determined accurately. For the interpretation
of our data it should be noted that streamers do not contribute to our statistics if they do not branch;
this is the case, e.g., at low pressure as in Fig.~\ref{drukzoom}~(c) or far from the electrodes as in
Fig.~\ref{vel}~(c). In our N$_2$, the branches are often very short, cf.~Fig.~\ref{n2airsimpel}~(b). As a
blurry surrounding, scattered light or accidentally overexposed pixels could be misinterpreted as a short
branch, only streamers with sharp, bright branches are evaluated in our N$_2$ which limits our data set.

\begin{figure}
\begin{center}
\vspace{0.5cm}
\begin{minipage}[b]{\linewidth}
\begin{minipage}[b]{.48\linewidth}
\includegraphics[height=6cm]{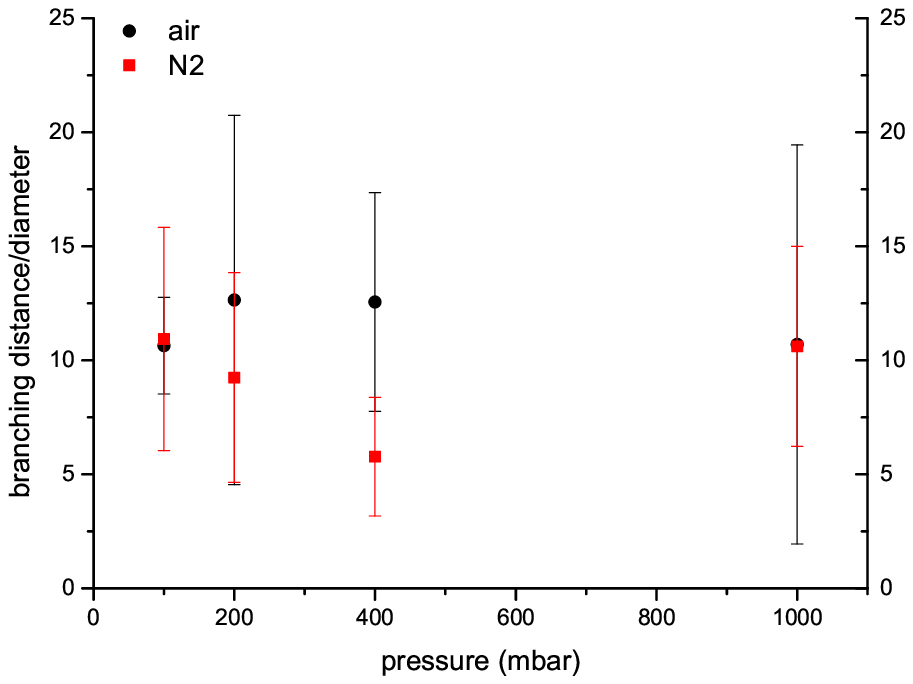}
\centering (a)
\end{minipage}\hfill
\begin{minipage}[b]{.48\linewidth}
\includegraphics[height=6cm]{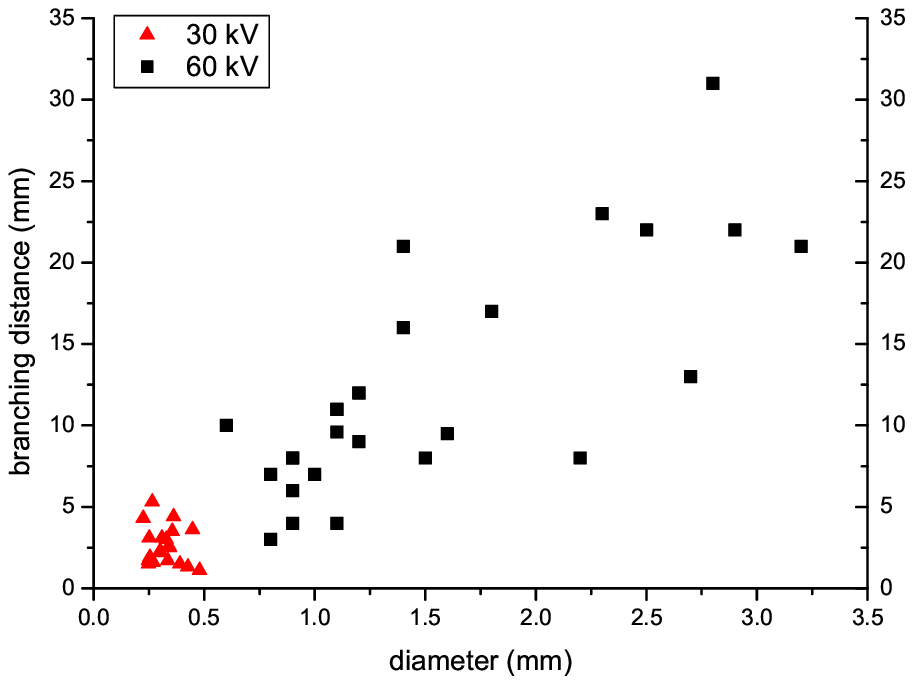}
\centering (b)
\end{minipage}\hfill
\end{minipage}
\caption{a) The ratio $D/d$ of branching lengths $D$ over streamer diameters $d$ at different pressures in a
4 cm gap just above the respective inception voltage. The points indicate the mean value, the bars indicate
the range of the statistical distribution; it must not be regarded as an error bar. b) The branching length
$D$ as a function of the streamer diameter $d$ in an 8 cm gap in air at 1 bar and different voltages. Here
data from \cite{bri06} with fast voltage rise time and thick streamers are plotted.}\label{n2airbranch}
\end{center}
\end{figure}

Figure \ref{n2airbranch}~(a) shows the branching ratio $D/d$ for several pressures and for voltages just
above the inception voltage. The streamer diameters in this case are small but not completely minimal. The
symbols in the figure (black dots for air and red squares for our N$_2$) indicate the median of the data.
The bars indicate the range from the lowest to the highest value and therefore the statistical distribution
of the branching lengths $D$. That there is actually a statistical distribution can be seen, e.g., in figure
\ref{figdiag}~(b) where the channel indicated by $D$ branches only once while the streamer on the right
branches two times over a shorter distance. If streamers at different pressures and otherwise similar
parameters are similar, then the length ratio $D/d$ does not depend on pressure. Within the scatter of the
data, this is indeed the case; a horizontal line $D/d = 11 \pm 4$ fits the data for air, and $9\pm 3$ fits
the data for our N$_2$. Here the error is calculated as the standard deviation of the points in the figure.

Figure \ref{n2airbranch}~(b) presents data for thicker streamers in air at 1 bar. Here the experimental
results from \cite{bri06} are evaluated. As in these experiments voltages of up to 60 kV were applied, and as
the voltage rise time was as short as 25 ns, streamers with diameters in the range from 0.2 to 3.25 mm were
created. The figure shows the branching length $D$ as a function of the streamer diameter $d$. The branching
length increases with the streamer diameter; it can be fitted with the line $D= (8 \pm 4)\cdot d$. For
comparison, the streamers of minimal diameter in air at 1 bar are formed at a voltage of about 10 kV; their
branching ratio has the larger value $D/d=11$ as shown in panel (a) of the figure. We conclude that at least
in air at 1 bar the absolute length $D$ between branches increases with diameter $d$ and voltage $U$, but
that the ratio $D/d$ of branching length over streamer diameter slightly decreases with diameter and voltage.


\section{Discussion and conclusions}\label{concl}

\subsection{Streamer diameter}

The primary aim of this paper is to test the similarity laws on streamers at different pressures $p$, or
more precisely, for streamers at different densities $n$, where $n=p/(k_BT)$ according to the ideal gas law.
For the streamer diameter, we have to analyze its reduced value $p\cdot d/T$. However, even at fixed
pressure, the streamer diameter can vary by more than an order of magnitude, depending on applied voltage
and voltage rise time~\cite{bri06}. Therefore we concentrate on the minimal diameter $d_{min}$. For the
reduced minimal streamer diameter in air, we find in our experiments the constant value $p\cdot
d_{min}=0.20\pm 0.02~{\rm mm\cdot bar}$ at room temperature when pressure changes from 0.013 to 1 bar. This
value agrees with old measurements of repetitive positive streamers in ambient air with DC voltage yielding
diameters of the order of $0.2\sqrt{\log 2}~{\rm mm} \approx 0.17 {\rm mm}$ for the full width at half
maximum (FWHM) \cite{Gibert}; the DC drive is an extreme case of slow voltage rise; consistently with our
findings, it creates streamers of minimal diameter.

It is quite unexpected that the reduced streamer diameters in air in this pressure range turn out to be
constant. As discussed in section~\ref{Townsend}, deviations from the similarity law are expected above
$p\approx 40$~mbar, because the excited nitrogen states that are responsible for photo-ionization,
increasingly loose their energy through collisions with neutral molecules rather than through
photo-emission.

It is also remarkable that the smallest reduced diameter of a sprite channel obtained in the telescopic
images in~\cite{gerken} is $p\cdot d/T=0.3\pm 0.2~\rm mm\cdot bar/(293 K)$. It is consistent with our
laboratory value even though the pressure was as low as $10^{-5}$~bar, cf.~Fig.~\ref{n2airEddie}. Of course,
it is not clear whether sprites of minimal diameter were included in the limited data set of~\cite{gerken};
and the higher background ionization at sprite altitude might lead to corrections to the similarity laws.

In the recent literature, the diameters of positive streamers in air are all within the range from 0.2 to 3
mm$\cdot$bar that was discussed in~\cite{bri06}, but no streamers of minimal diameter were reported
elsewhere. Pancheshnyi {\it et al.}~\cite{panchpre} report measurements and simulations with $p\cdot d =
0.5$ mm$\cdot$bar at 1 bar and $E=7$~kV/(cm$\cdot$bar), and with $p\cdot d=1.2$ mm$\cdot$bar at 0.5 bar and
$E=14$~kV/(cm$\cdot$bar). A stagnating streamer had a diameter of 0.4~mm at 1~bar in~\cite{PanStag} while
the same authors find streamer diameters between 1 and 3 mm at 1 bar in stronger fields in~\cite{Pan13cm},
both in experiments and in simulations. Liu and Pasko~\cite{pasko2006} report simulations with $p\cdot d$ =
1.0, 1.4 and 1.2 mm$\cdot$bar at 1, 0.014 and $6\cdot10^{-5}$ bar and $E=5$~kV/(cm$\cdot$bar).

For our N$_2$ with an impurity level of about 0.1 \%, we find a reduced minimal diameter of $p\cdot
d_{min}=0.12\pm 0.03~{\rm mm\cdot bar}$ at room temperature, but the spread of the data is larger than in
air. This is probably due to temporal variations of the impurity concentration; a new experimental setup is
therefore now in development.

\subsection{Streamer velocity}

The similarity laws state that the velocities of streamers with the same reduced diameter are the same. We
concentrate on streamers with minimal diameter. Unfortunately, their velocity is very difficult to determine
at pressures below 1 bar. This is due to the increasing jitter of the inception time which makes appropriate
timing of the iCCD photography increasingly difficult. A second complication is that it is difficult to
ascertain that the streamer segment ends because the exposure time of the camera finishes, and not because
the streamer stops by itself. The minimal velocity at 1 bar in air is 0.1 mm/ns = $10^5$ m/s, the higher
values at lower pressure in figure~\ref{n2airvel} do not belong to streamers of minimal diameter. It is
remarkable that the lowest sprite velocity reported in \cite{moudry} is $10^5$ m/s as well. However, this
lower limit is reported only occasionally for sprites while values of up to $10^8$ m/s are observed
\cite{McHarg02}. Now with fast voltage rise time, streamer diameters and velocities also increase to 3 mm
and $4\cdot 10^6$ m/s for a voltage of 96 kV in ambient air~\cite{bri08jpd}; probably they increase further
with higher applied voltage. Therefore maximal velocities of streamers or sprites cannot be compared while
minimal velocities are well defined and similar.

The streamer velocities in our N$_2$ are quite different. At 1 bar they are much lower, down to $2\cdot10^4$
m/s, but at 0.4 and 0.065 bar the values are quite the same or even higher than in air. Again we might not
observe minimal streamers at the lower pressures, and the influence of the impurities again is uncertain and
might increase with decreasing pressure. As positive streamers can only propagate due to photo-ionization,
due to electron detachment from O$_2^-$ or due to background ionization, and as the photo-ionization rate
decreases and the photo-ionization lengths increase with decreasing oxygen concentration~\cite{Luque08},
quite different velocities can be expected in our N$_2$. Whether positive streamers in pure nitrogen without
background ionization propagate at all, is not known. The only other data is presented in~\cite{yi02}, it is
not conclusive.

\subsection{Size of the initial ionization cloud}

The discharge starts at the anode point as an oblate cloud with an aspect ratio of height over width
$h/w=0.7$. This cloud is clearly observed in air at pressures of 0.4 bar and lower. At 1 bar the cloud is
difficult to distinguish as it is very small, and the iCCD images are often overexposed at the tip. The
cloud size at the inception voltage approximately obeys a similarity law; the reduced cloud height is
approximately $h\cdot p = 0.5 \pm 0.2 ~\rm mm\cdot bar$. In our N$_2$, the cloud is less pronounced.

\subsection{Branching structure}

Elsewhere we have measured the distribution of branching angles~\cite{Nijdam08}. Here we have introduced and
measured the branching ratio $D/d$ to characterize the shape of the streamer tree; here $D$ is the distance
between branching events and $d$ the streamer diameter just before branching. In air, this ratio is $D/d =
11 \pm 4$ for streamers of minimal diameter for all pressures in the range from 0.065 to 1 bar. Therefore
the overall branching pattern is quite independent of pressure, even though the stochastic fluctuations are
larger at higher pressures as discussed in section~\ref{Townsend}. If random ionization avalanches would
play a major role in the branching process (cf.~the discussion in section 5 of~\cite{ebe06}), we would
expect relatively more branching, i.e., a smaller $D/d$, at higher pressures.

For thick streamers in air at 1 bar, we find $D/d$ = 8 $\pm$ 4. Therefore for fixed pressure, thicker
streamers are longer from one splitting to the next, but measured on the scale of their diameter, they
branch slightly more frequently. This trend agrees with observations of Yi and Williams \cite{yi02}; they
report an average branching distance of 3 mm for positive streamers with a diameter of about 2 mm in N$_2$
at 1 bar at 98 kV. This yields $D/d = 1.5$ in qualitative agreement with our observation that the ratio
$D/d$ decreases with increasing voltage at a fixed pressure. It also indicates that streamers in nitrogen
branch more frequently than in air.

The branching ratio in our N$_2$ is $D/d = 9 \pm 3$ for thin streamers in the same pressure range. This
value is somewhat lower than in air. As also the diameter $d$ of N$_2$ streamers is smaller, the absolute
streamer length $D$ between branching is only half of that in air at the same pressure. This can be seen
qualitatively in figure \ref{n2airsimpel}. For a further comparison of streamers in air and in our N$_2$, we
refer to the \ref{app}.

\subsection{Outlook}

We conclude that the reduced diameter $p\cdot d/T$ of minimal streamers is constant for pressures in the
range from $10^{-5}$ to 1 bar at room temperature, while present theory predicts a change of behavior above
approximately 0.04 bar. The velocity, initial cloud size and branching structure also scale quite well with
pressure, but due to the experimental limitations this could only be tested in a less extended pressure
range. Evaluating the results in our N$_2$ is somewhat problematic because presently the impurity level in
the gas is not sufficiently controlled. In the near future we plan new experiments to improve this
situation. In any case, the comparison of our experimental results with theory at varying pressure and gas
composition and the discussion of the microscopic physical mechanisms will pose challenges for the future.

\ack This work was supported by STW under contract number 06501 and by NWO under contract number 047.016.017.


\appendix
\section{Comparison of streamers in air and nitrogen}\label{app}

\begin{figure}
\begin{center}
\begin{minipage}[b]{.48\linewidth}
\centering
\includegraphics[height=6cm]{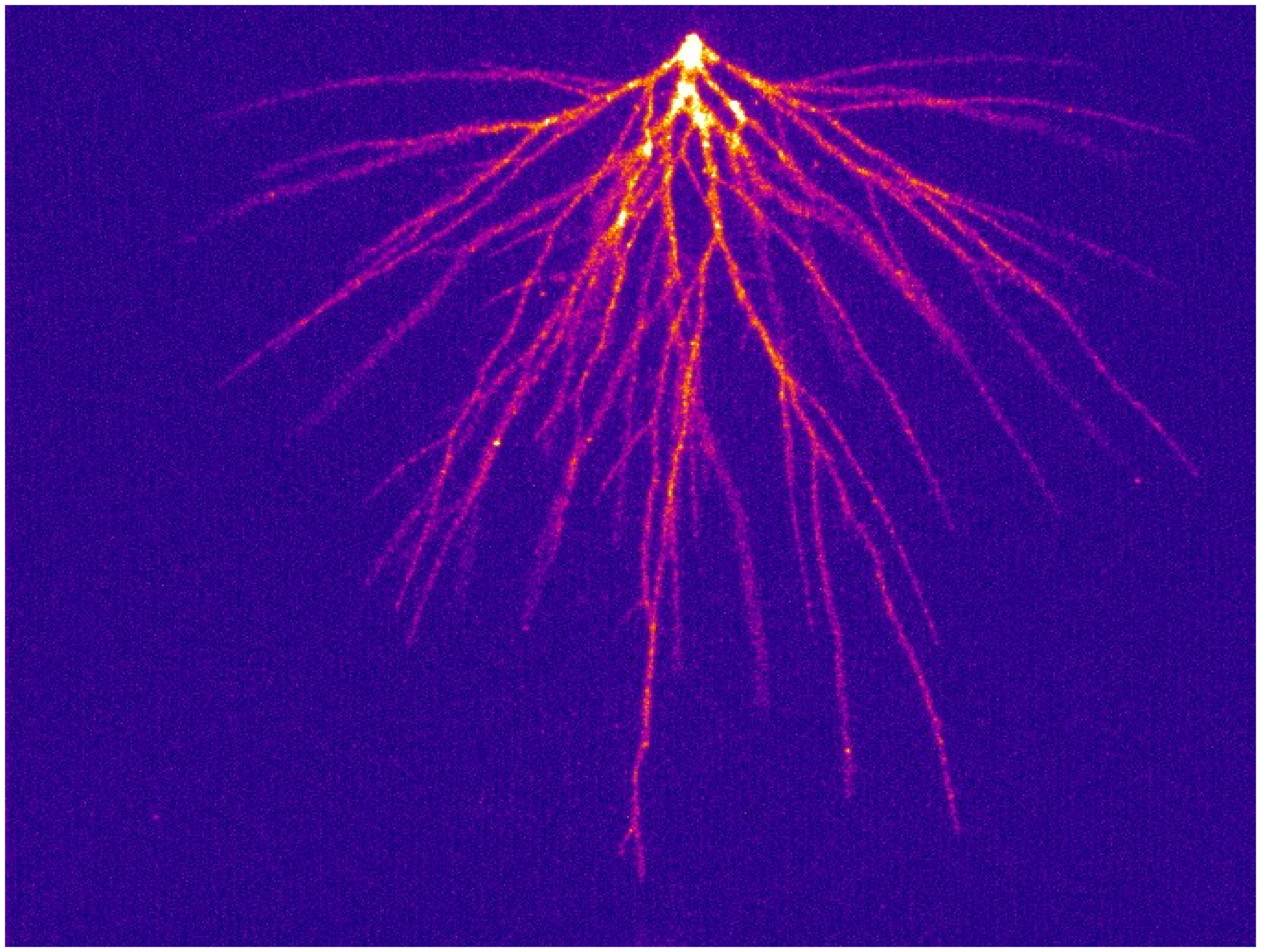}
a) air, 1 bar, 21 kV \\~\\
\includegraphics[height=6cm]{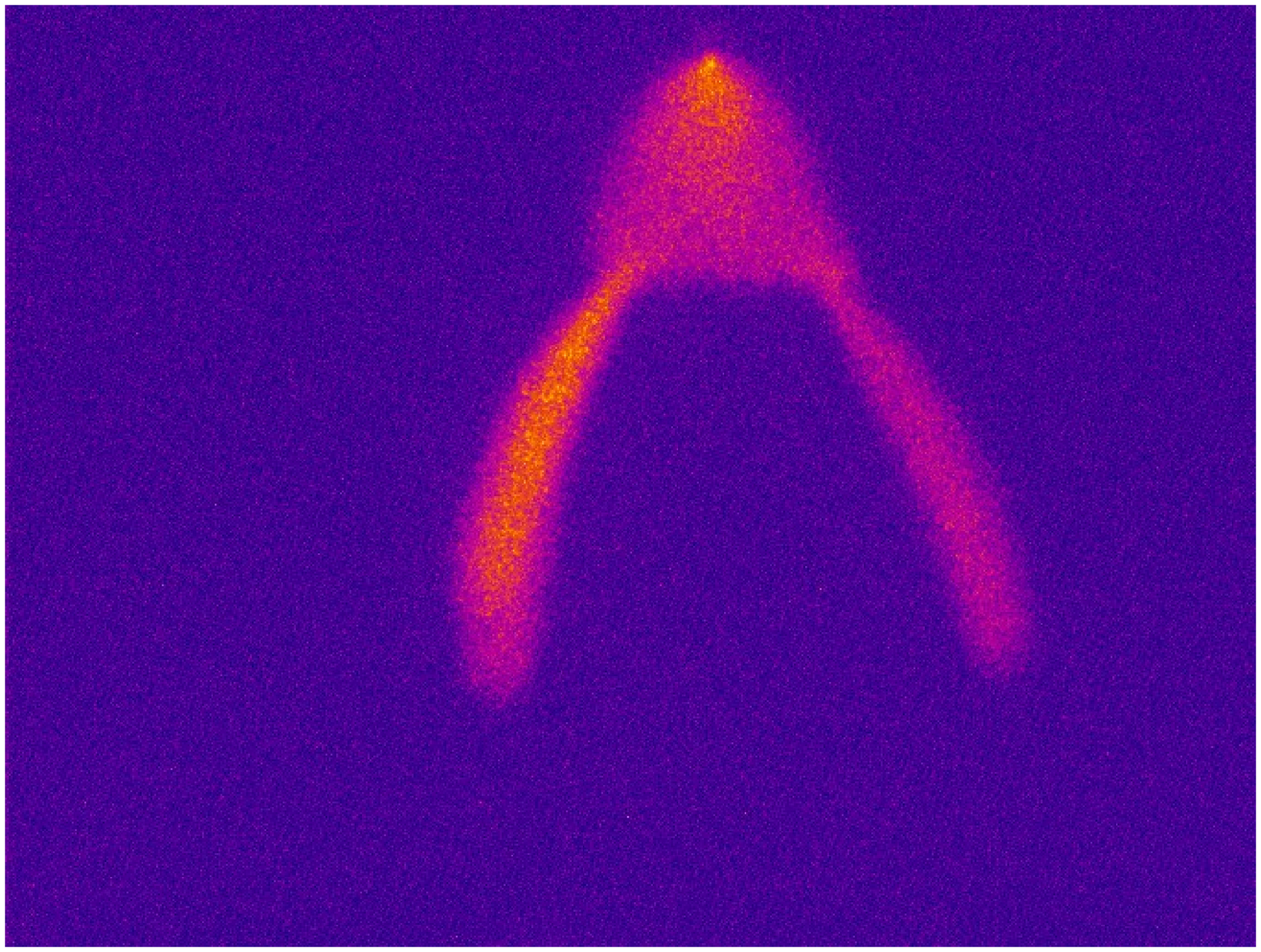}
c) air, 0.1 bar, 5 kV
\end{minipage}\hfill
\begin{minipage}[b]{.48\linewidth}
\centering
\includegraphics[height=6cm]{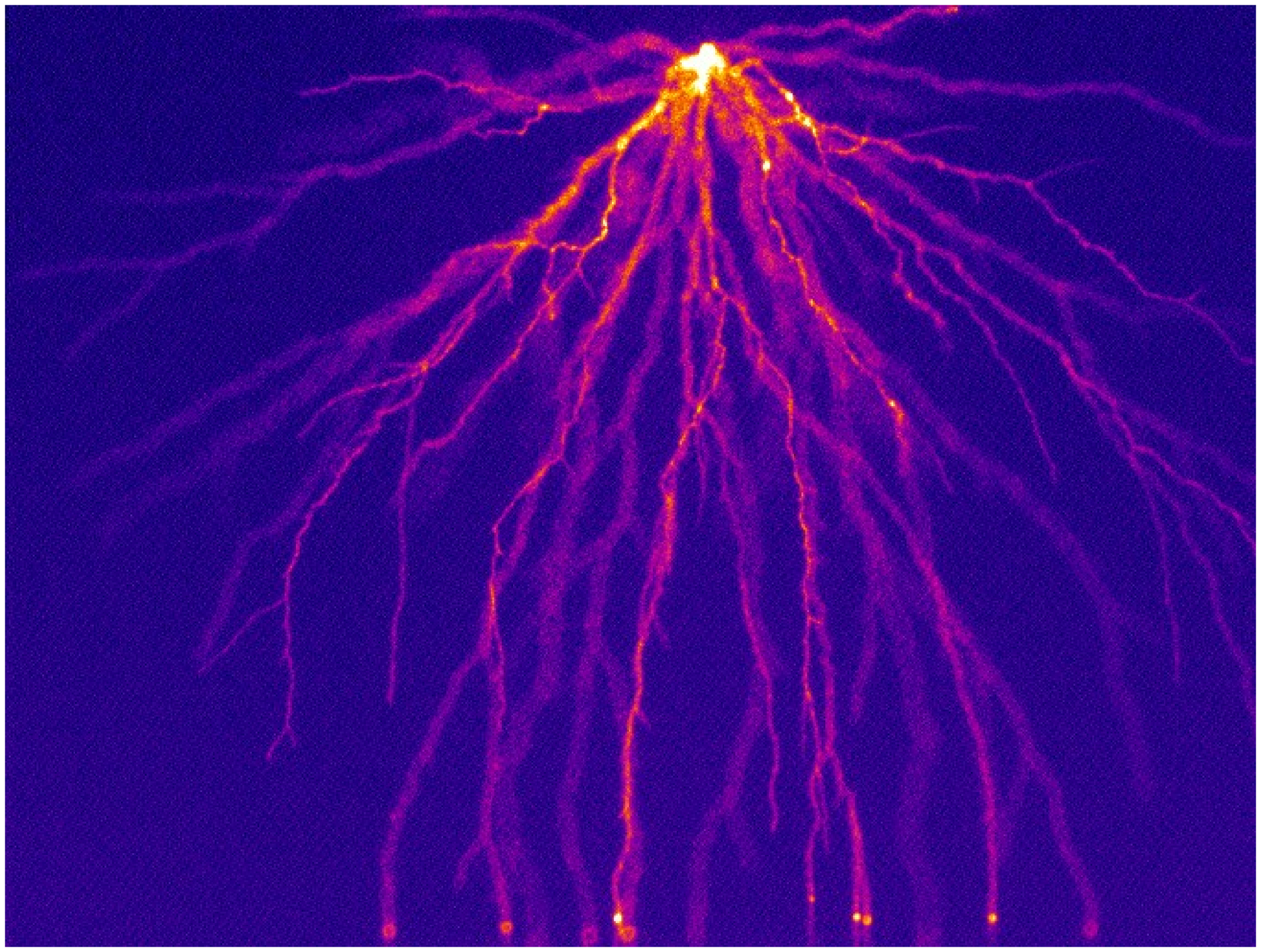}
b) N$_2$, 1 bar, 19 kV \\~\\
\includegraphics[height=6cm]{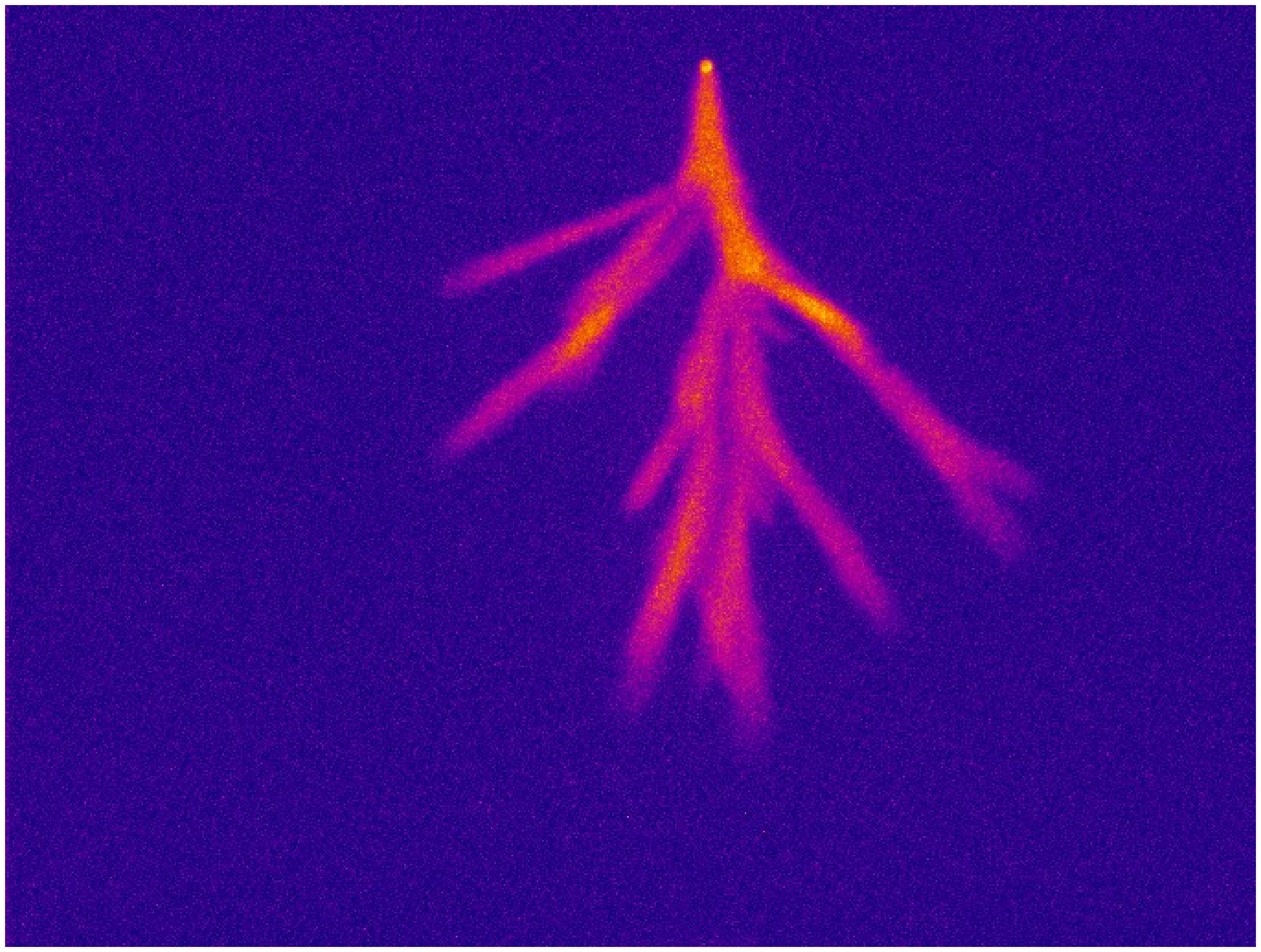}
d) N$_2$, 0.1 bar, 5 kV
\end{minipage}
\caption{Streamers in a 4 cm gap in air (left column) and in our N$_2$ (right column) at $p=1$~bar and
$U\approx 20$~kV ($p\cdot d_{gap}=40$~mm$\cdot$bar, upper row) and at $p=0.1$~bar and $U=5$~kV ($p\cdot
d_{gap}=4$~mm$\cdot$bar, lower row). The camera delay is $T=0$, the exposure time $\Delta t$ and pulse
frequency are: a) 0.6 $\mu$s, single shot; b) 1.5 $\mu$s, 1 Hz;  c) 0.7 $\mu$s, 1 Hz; and d) 0.53 $\mu$s, 10
Hz. A part of the streamers eventually reaches the plate electrode, but only after the exposure times of
photos (a), (c) and (d).}\label{n2airuiteinde}
\end{center}
\end{figure}

Though nitrogen is the major component of air, streamers in nitrogen and air have two important distinctions
on the microscopic level: $(1)$ Nitrogen-oxygen mixtures have an efficient photo-ionization mechanism that
allows streamers to propagate through a nonlocal ionization mechanism against the electron drift direction
(though the model parameters for this interaction are not very well known); and $(2)$ free electrons are
easily lost in air through attachment to oxygen. Therefore, it is quite instructive to compare discharges in
these two gases. Note that a quantitative link between these microscopic mechanisms and the macroscopic
appearance is emerging only now, except for stating that electron attachment limits the duration of
discharges in air.

The differences between streamers in air and nitrogen are demonstrated on typical photographs taken in the
gap of 40 and 160 mm. Another good example of long streamers in the 160 mm gap is given in ~\cite{bri08}. We
recall that our N$_2$ contains about 0.1 \% oxygen and our nitrogen-oxygen mixture 0.2 \% oxygen; the oxygen
concentration in air is about 100 times higher than in our N$_2$ or in our mixture.

Figure~\ref{n2airuiteinde} shows photographs of streamers in air and in our N$_2$ at pressures of 1 and 0.1
bar. The comparison of panels (a) and (b) shows clearly that streamers in nitrogen move more in a zigzag
manner while the streamers in air are straighter. Furthermore, the streamers in nitrogen branch more
frequently. The panels also show that streamers in nitrogen have a better contrast between in- and
out-of-focus structures. Panels (c) and (d) show that the streamers in N$_2$ are thinner, and that the light
emitting cloud at the electrode tip is clearly visible in air while it is less pronounced in nitrogen. The
camera timing in panels (c) and (d) is such that the streamers are still propagating; with a longer exposure
time, one would see them reaching the cathode plate. Therefore, the different light emission structure of
the propagating streamer tips can be studied on these panels: The streamer tips in air in panel (c) are
round and stop abruptly, while the tips in N$_2$ in panel (d) are more diffuse. The length along the
streamer over which the intensity drops to 50\% of its maximal value is about 4 times longer in N$_2$.

\begin{figure}
\begin{center}
\begin{minipage}[b]{.48\linewidth}
\includegraphics[height=6cm]{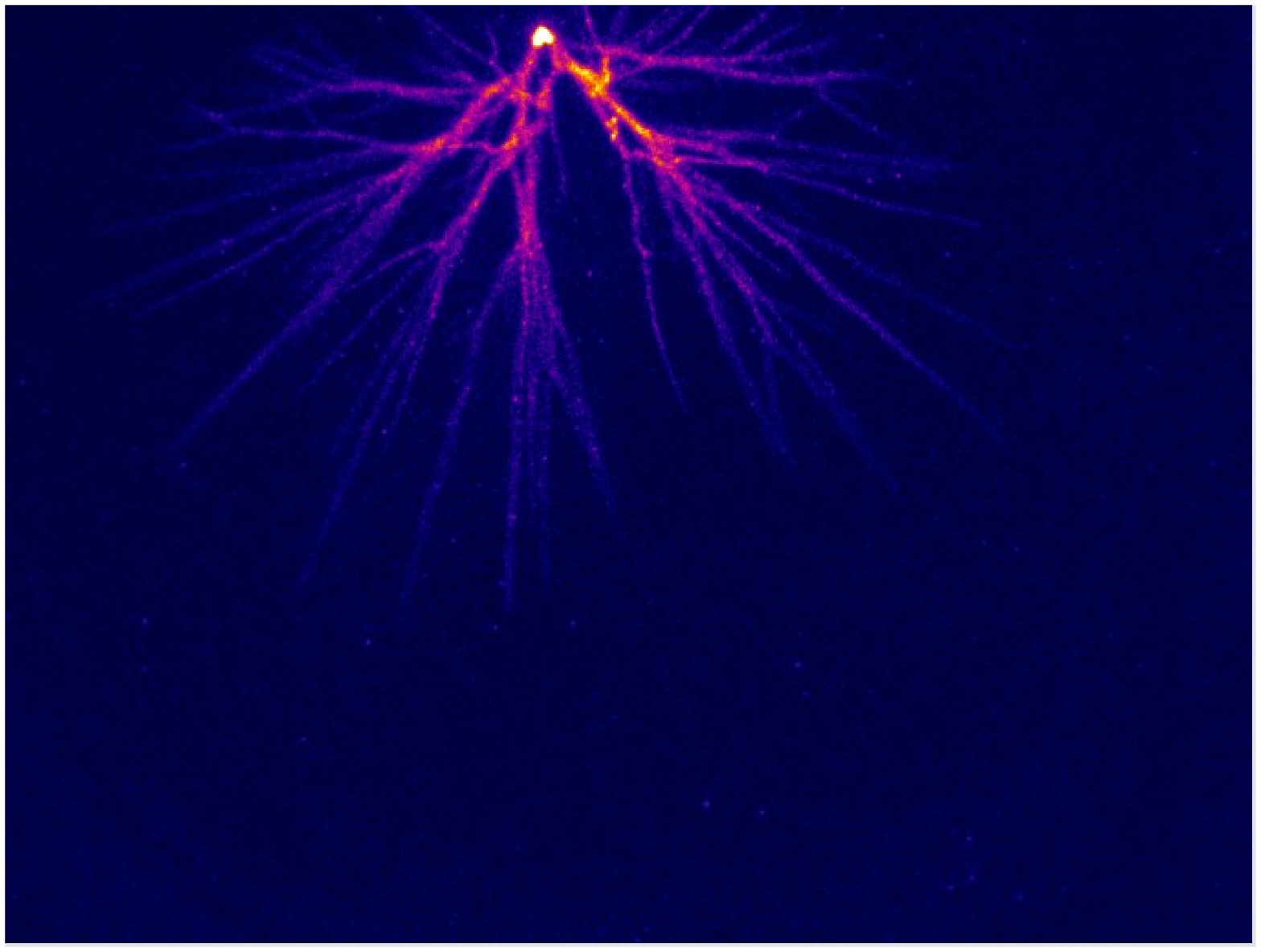}
\centering (a): air, oxygen concentration is 100 times higher than in the mixture (c)
\\~\\
\includegraphics[height=6cm]{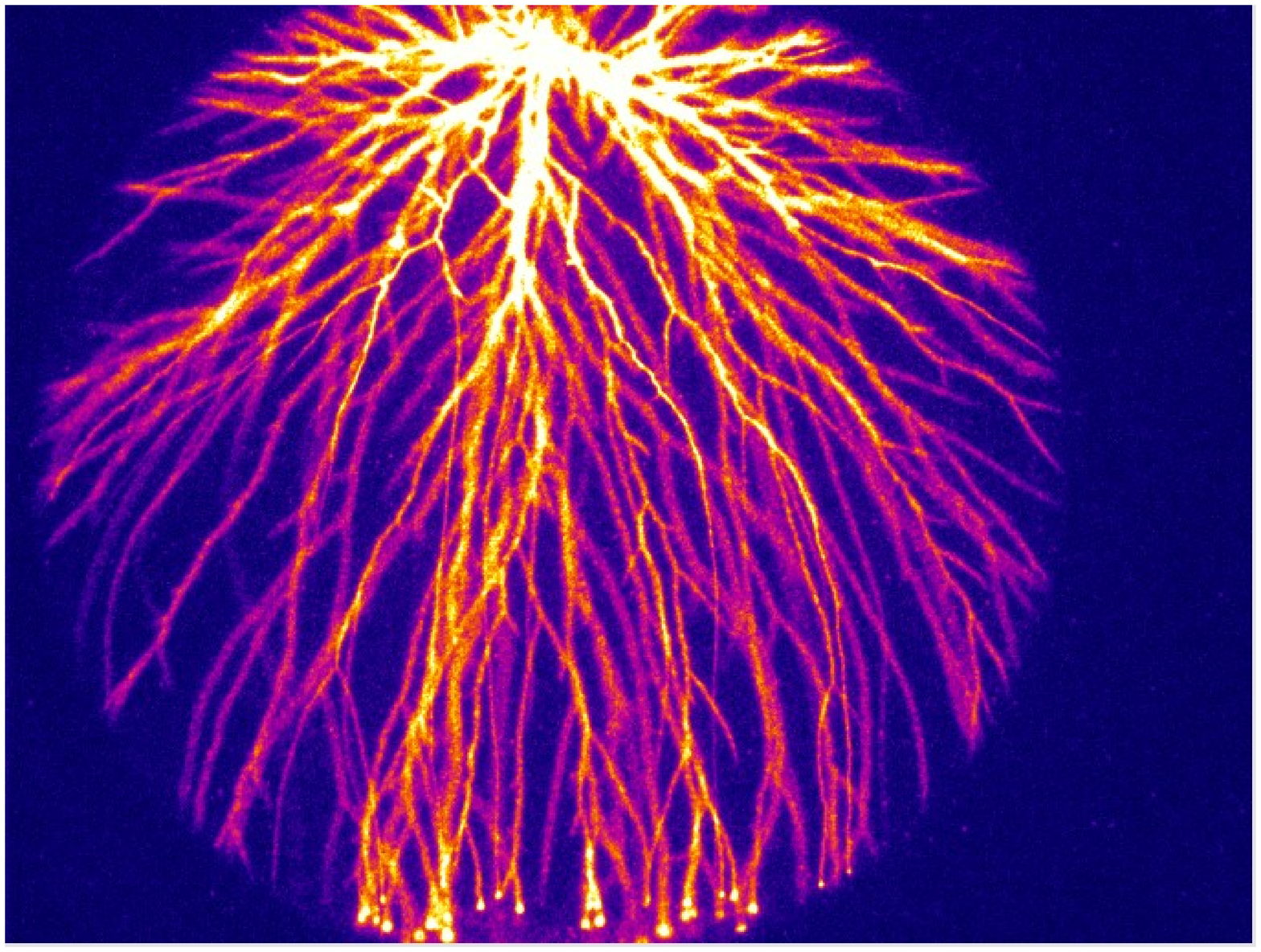}
\centering (c): nitrogen-oxygen mixture (99.8:0.2), \\
same view and intensity scale as in (a)
\end{minipage}\hfill
\begin{minipage}[b]{.48\linewidth}
\includegraphics[height=6cm]{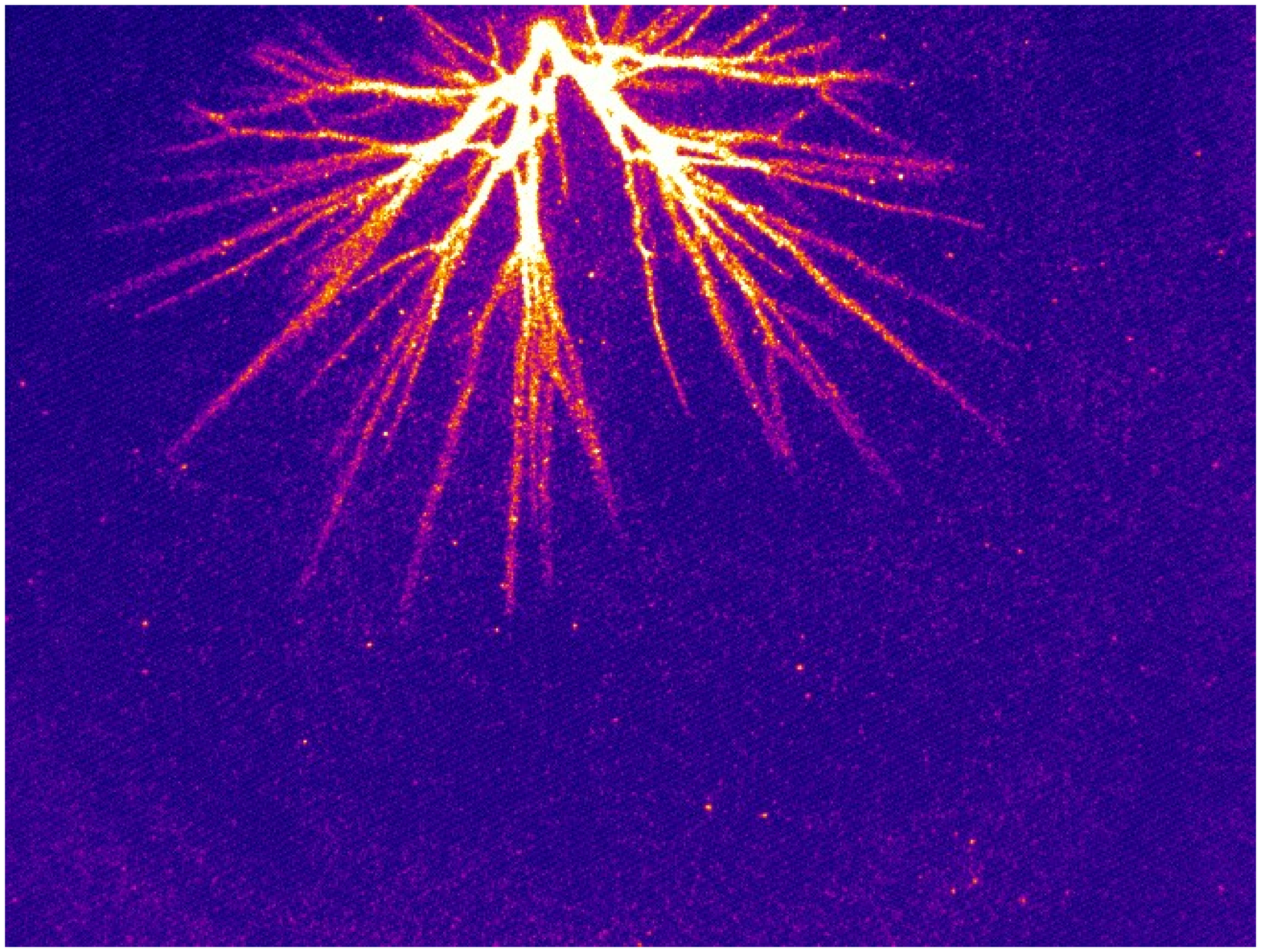}
\centering (b): the data in air of (a), \\ but with enhanced intensity scale
\\~\\
\includegraphics[height=6cm]{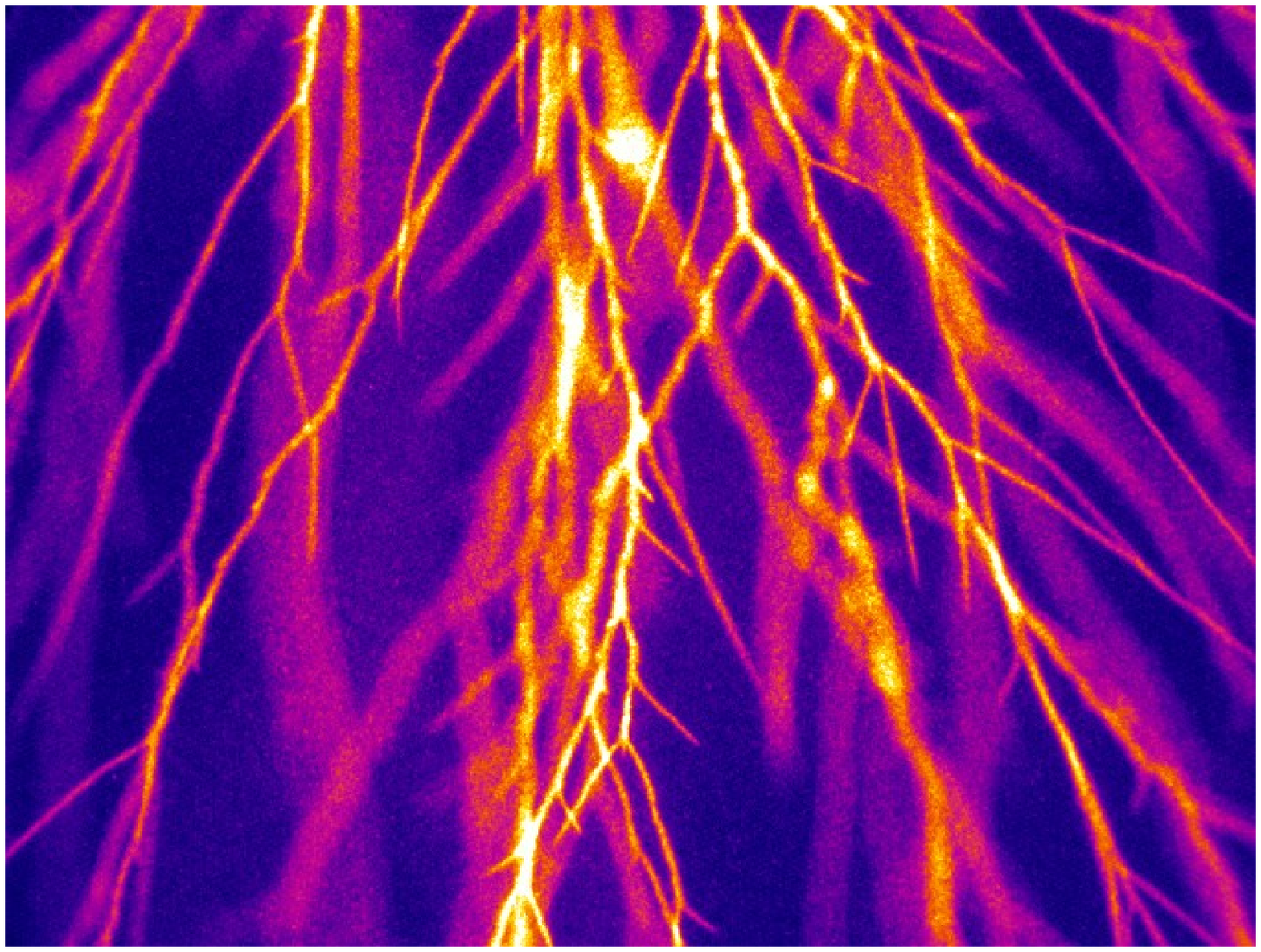}
\centering (d): same parameters as in (c), \\ but camera has moved closer by.
\end{minipage}
\caption{Streamers in a 16 cm gap at $p=0.4$ bar and $U\approx21$ kV ($p\cdot d_{gap}=64$~mm$\cdot$bar);
(a,b): air at 20 kV, and (c,d): nitrogen-oxygen mixture (99.8 : 0.2) at 22 kV. The oxygen concentration in
air is about 100 times higher than in the mixture. Panels (a) and (c) are on the same intensity scale, (b)
has a different scale to show the complete streamer structure. Panel (d) zooms into the region of 65 to 110
mm from the needle tip. The figures show the full evolution, hence the streamers in air do not cross the
gap.}\label{n2airvergelijk}
\end{center}
\end{figure}

Figure \ref{n2airvergelijk} shows streamers in air and in the nitrogen-oxygen mixture with ratio 99.8:0.2
with exactly the same setup and camera settings apart from the gas composition. (The circular outer edge of
the streamers visible in the figure is the edge of the quartz window in the vacuum vessel that limits the
view when imaging the 16 cm gap, cf.~Fig.~\ref{setup}.) The exposure time is 70 $\mu$s, which is longer than
the voltage pulse duration, therefore the whole path of the streamers towards the cathode plate can be seen.
Panels (a) and (c) show streamers in air and in the mixture with identical intensity coding; the streamers in
the mixture are much more intense, and they bridge the complete gap. The rather faint data of panel (a) is
reproduced with a more sensitive color coding in panel (b); here it can be verified that the streamers in air
do not bridge the gap. The intensity difference between air and the mixture is not related to the bridging of
the gap, since it is also present at other pressures and voltages where the streamers in the mixture do not
bridge the gap. Panels (b) and (c) show that the streamers in the mixture branch more. Note that the results
in our N$_2$ are similar. Panel (d) zooms into the region of 65 to 110 mm below the tip (i.e. halfway the
gap) for the streamers in the mixture. It shows many branches that mostly remain short. We remark that we use
figures such as panel (d) to measure diameters in gaps of 16 cm, since the resolution of panel (c) is
insufficient and leads to measuring artifacts~\cite{bri06}.


\end{document}